\documentclass[nofootinbib,twocolumn,aps,superscriptaddress,longbibliography]{revtex4-2}
\usepackage{amsmath}
\usepackage{amssymb}
\usepackage{todonotes}
\usepackage[unicode=true,colorlinks=true,citecolor=blue,urlcolor=blue]{hyperref}
\usepackage{float}
\usepackage{bm}
\usepackage{epsfig}
\usepackage{lipsum}
\usepackage{xcolor,soul}
\usepackage[normalem]{ulem}
\usepackage{physics}
\usepackage{comment}
\usepackage{balance}
\renewcommand {\phi}{{\varphi}}

\usepackage{flushend}

\begin{document}

\title{Topological nature of edge states for one-dimensional systems without symmetry protection}
\affiliation{Department of Applied Physics, Stanford University, Stanford, California 94305, USA}
\affiliation{Department of Electrical Engineering, Ginzton Laboratory, Stanford University, Stanford, California 94305, USA}
\affiliation{Department of Physics of Complex Systems, Weizmann Institute of Science, Rehovot 7610001, Israel
\\ $^\dagger$These authors contributed equally to this work.}
\author{Janet Zhong$^\dagger$}
\affiliation{Department of Applied Physics, Stanford University, Stanford, California 94305, USA}
\author{Heming Wang$^\dagger$}
\affiliation{Department of Electrical Engineering, Ginzton Laboratory, Stanford University, Stanford, California 94305, USA}
\author{Alexander N. Poddubny}
\affiliation{Department of Physics of Complex Systems, Weizmann Institute of Science, Rehovot 7610001, Israel
\\ $^\dagger$These authors contributed equally to this work.}
\author{Shanhui Fan}
\email{shanhui@stanford.edu}
\affiliation{Department of Applied Physics, Stanford University, Stanford, California 94305, USA}
\affiliation{Department of Electrical Engineering, Ginzton Laboratory, Stanford University, Stanford, California 94305, USA}

\begin{abstract}
We numerically verify and analytically prove a winding number invariant that correctly predicts the number of edge states in one-dimensional, nearest-neighbor (between unit cells), two-band models with any complex couplings and open boundaries. Our winding number uses analytical continuation of the wave-vector into the complex plane and involves two special points on the full Riemann surface band structure that correspond to bulk eigenvector degeneracies. Our winding number is invariant under unitary or similarity transforms. We emphasize that the topological criteria we propose here differ from what is traditionally defined as a topological or trivial phase in symmetry-protected classification studies. It is a broader invariant for our model that supports non-zero energy edge states and its transition does not coincide with the gap closing condition. When the relevant symmetries are applied, our invariant reduces to well-known Hermitian and non-Hermitian symmetry-protected topological invariants.
\end{abstract}
\date{\today}
\maketitle

\textit{Introduction-- }There has been substantial interest in the topology of band structures~\cite{hasan2019colloquium,chiu2016classification,cayssol2021topological}. In Hermitian systems, the bulk-edge correspondence indicates the connection between topological invariants defined using the bulk states and the existence of edge states~\cite{alase2023wiener,chiu2016classification}. In one dimension, this correspondence is established for systems with certain symmetries as classified under the ten Altland-Zirnbauer symmetry classes, known as the 10-fold way~\cite{altland1997nonstandard,kitaev2009periodic, ryu2010topological, chiu2016classification}. In non-Hermitian systems, topological invariants for zero-energy edge states have also been established~\cite{yao2018edge,yokomizo2019nonbloch,ashida2020nonhermitian,lee2019anatomy,kunst2018biorthogonal,bergholtz2021exceptional,esaki2011edge,ding2022nonhermitian,lin2023topological}. The required symmetries are classified by the 38-fold way~\cite{kawabata2019symmetry,zhou2019periodic}. 

In this Letter, we focus on asymmetric one-dimensional Hermitian and non-Hermitian models. Such an asymmetric model has periodicity and hence translational symmetry in the bulk, but has no other symmetries (we do not consider random disorders that may break translational
symmetry). Although asymmetric models have no symmetry-protected topological invariants, they may still support edge states and moreover these edge states may persist over a range of parameters of the Hamiltonians (see Supplementary Material (SM) Sec.~\ref{supp:nosymmetry}~\cite{supplemental}). The observations above motivates us to look for a more general topological invariant that accounts for the existence of the edge states in asymmetric models.
\begin{figure}[t!]
\centering
\includegraphics[width=0.5\textwidth]{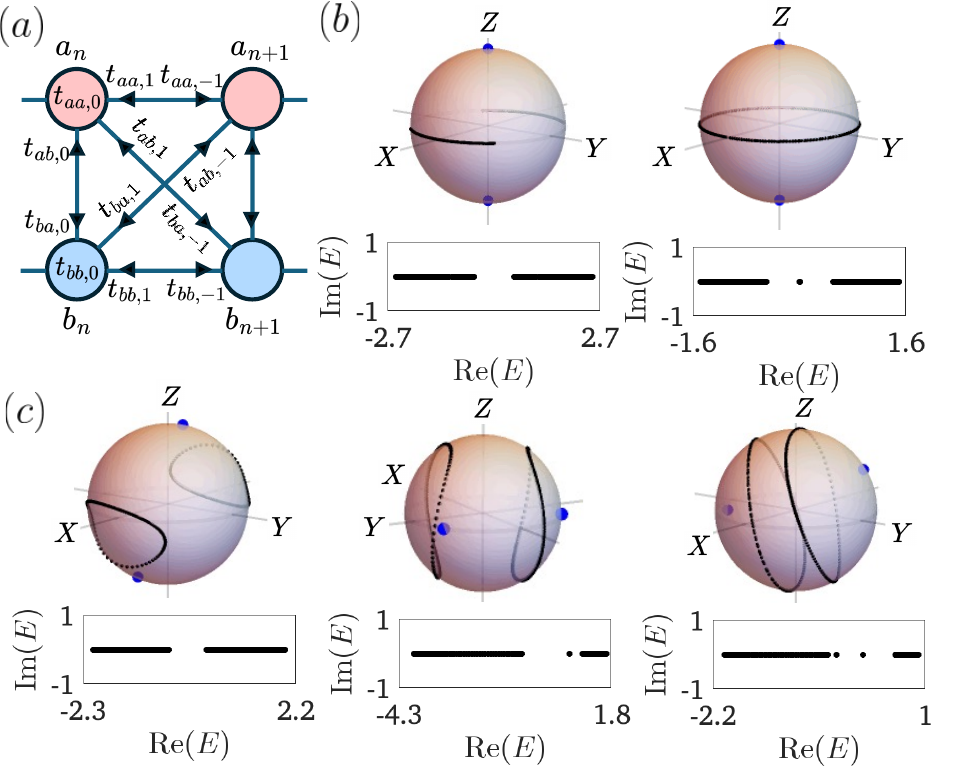}
\caption{(a) Two unit cells of the model in Eq.~\eqref{Eq:H}. (b) Hermitian SSH edge-state invariant on $M$-Riemann sphere. (c) Hermitian random model and invariant on $M$-Riemann sphere corresponding to 0, 1 and 2 edge states. Model parameters are provided in the End Matter. Here, the two $M_{\text{deg}}$ points are blue dots and the black loops are the image of the Brillouin zone on the $M$-Riemann sphere. }
\label{fig:couplings}
\end{figure}

In this Letter, we consider a class of two-band models with complex couplings between nearest-neighbor unit cells. This class of models have been extensively studied in the topological physics literature~\cite{alvarez2018non,liang2022topologial,han2021topological,lieu208topological,halder2024circuit,halder2023properties,kachin2024ultra,yuce2015topological,verma2024non,wang2023sub,kawarabayashi2021bulk,saha2023topological,su1979solitons,lee2016anomalous,lee2019anatomy,yokomizo2019nonbloch,borgnia2020nonhermitian,longhi2019probing,kunst2018biorthogonal,guo2021nonhermitian,mccann2023catalog,mccann2023catalog,hou2023topological,chiu2016classification,lai2025non}. We introduce a topological invariant for the existence of edge states in such models when an open boundary condition is imposed. This invariant builds entirely from properties of the bulk band structure and the bulk eigenstates, as described in the language of Riemann surfaces. It naturally reduces to known invariants in systems with symmetry and generalizes these known invariants to non-Hermitian systems without the need of symmetry constraints. The invariant takes a simplified form in Hermitian systems. We also show that the removal and emergence of edge states under the open boundary condition (OBC) in asymmetric models is not connected to the gap condition, and our invariant provides a measure of the strength of perturbation that is required in order to create or eliminate an edge state (see SM Sec.~\ref{supp:robustness}).

\textit{Model -- }We consider the following one-dimensional, two-band model as depicted in Fig.~\ref{fig:couplings}(a):
\begin{equation}
H(z)=\left(\begin{array}{ll}
t_{a a, 0}+\frac{t_{a a,-1}}{z}+t_{a a,1} z & t_{ab, 0}+\frac{t_{ab,-1}}{z}+t_{ab,1} z \\
t_{ba, 0}+\frac{t_{ba,-1}}{z}+t_{ba,1} z & t_{b b, 0}+\frac{t_{b b,-1}}{z}+t_{b b,1} z
\end{array}\right).
\label{Eq:H}
\end{equation}
In this model, each unit cell contains an $a$ and $b$ site. $t_{\mu \nu,n}$ represents the coupling from the $\nu$ site in the $(m+n)$-th unit cell to the $\mu$ site in the $m$-th unit cell ($t_{aa,0}$ and $t_{bb,0}$ are on-site potentials)~\cite{wang2024nonhermitian}. Here $z=e^{ik}$ is the phase factor, where $k$ is the wave-vector and is in general complex~\cite{yao2018edge,alase2017generalization}. This is the most general model containing only nearest neighbor coupling between unit cells. The open boundary condition (OBC) is when $(\psi_{na},\psi_{nb})^T=(0,0)^T$ for $n=0$ and $n=N+1$ where $\psi_{na},\psi_{nb}$ are amplitudes at the sublattice sites within the 
$n$-th unit cell and $N$ is the number of unit cells in a finite chain. In the $N \rightarrow \infty$ limit, the majority of the OBC eigenvalues form a continuum OBC band consisting of a dense set of energies on the complex $E$-plane (or the $\text{Re}(E)$ axis for Hermitian models)~\cite{yokomizo2019nonbloch}. We call these OBC states the `bulk states.' However, some OBC eigenvalues may appear as discrete, isolated eigenvalues at an energy away from these continuum OBC bands. We call the states with these discrete OBC eigenvalues the `edge states.' For Hermitian models, edge states are localized whereas bulk states are extended. For non-Hermitian models of Eq.~\eqref{Eq:H}, bulk or edge states may both be localized~\cite{bergholtz2021exceptional,lee2016anomalous,yao2018edge,okuma2020topological,zhang2020correspondence}. We are interested in the topological origin of these edge states in asymmetric models. For Hermitian models, in Eq.~\eqref{Eq:H}, we would have $t_{\mu \nu, n}=t_{\nu \mu, -n}^*$. Eq.~\eqref{Eq:H} can also be written as:
\begin{equation}
H(z)=\sum_{i=0, x, y, z} d_i(z) \sigma_i,
\end{equation}
where $\sigma_i$ are Pauli matrices (and $\sigma_0=\mathbb{I}$). The $d_i$'s are in general complex (real) for non-Hermitian (Hermitian) systems. While $d_0(z)$ is sometimes omitted in previous studies, this term is quite important in non-Hermitian systems as its $z$ dependence can change the eigenvalue braiding topology~\cite{fu2024braiding,li2022topological,wojcik2020homotopy,hu2021knots,li2021homotopical,yang2024homotopy}.

The right eigenvalue equation of Eq.~\eqref{Eq:H} reads
\begin{equation}
\left[\begin{array}{cc}
d_0(z)+d_z(z) & d_x(z)-i d_y(z) \\
d_x(z)+i d_y(z) & d_0(z)-d_z(z)
\end{array}\right]\left[\begin{array}{l}
a(z)\\
b(z)
\end{array}\right]=E(z)\left[\begin{array}{l}
a(z) \\
b(z)
\end{array}\right],
\label{Eq:blochequation}
\end{equation}
where $E(z)$ is given by the characteristic polynomial
\begin{equation}
    \operatorname{det}\left[H(z)-E(z) \mathbb{I}\right]=0 .
    \label{Eq:charpoly}
\end{equation}
Eqs.~\eqref{Eq:H} to~\eqref{Eq:charpoly} is a result of the generalized Bloch theorem~\cite{alase2017generalization} which gives a bulk band structure that applies for {arbitrary boundary conditions so long as translational symmetry is preserved in the bulk~\cite{alase2017generalization,guo2021exact,wang2024nonhermitian} (see SM Sec.~\ref{supp:riemann}). The number of $z$ solutions at every $E$ in the characteristic polynomial is $p$, where $p$ is the order of the linear recurrence relations from the equations of motion~\cite{wang2024nonhermitian,bruna2013complex}. The general solution for these relations is the generalized Bloch ansatz $\binom{\psi_{na}}{\psi_{nb}}=\sum_{i=1}^pC_i z_i^n \binom{a_i}{b_i}$~\cite{alase2017generalization,zhang2020subradiant, yokomizo2020non,wang2024nonhermitian,hou2023topological,yao2018edge,wang2022amoeba}. This gives the real-space eigenvector at energy $E$ where $C_i$ is determined by the boundary conditions. A simple application of this is the OBC modes of a Hermitian Su-Schrieffer-Heeger (SSH) model where $p=2$. For bulk states, $|z|=1$ and we recover our typical Bloch ansatz. For edge states, $|z_1|<1$ and $|z_2|>1$ and the eigenstate is composed of both decaying and amplifying waves. For our model in Eq.~\eqref{Eq:H}, $p=4$, there are four solutions for $z$ at every $E$, which we label in order of magnitude: $|z_1| \leq |z_2| \leq |z_3| \leq |z_4|$. 

The study of topological edge states typically involves the integration of the Berry connection found from the bulk eigenvector of Eq.~\eqref{Eq:H} over the Brillouin zone (BZ) for Hermitian cases~\cite{chiu2016classification,asboth2015ashort}. For non-Hermitian cases, we integrate over the generalized Brillouin zone (GBZ)~\cite{yao2018edge,yokomizo2019nonbloch}. The GBZ for the model in  Eq.~\eqref{Eq:H} are the solutions for Eq.~\eqref{Eq:charpoly} when $|z_2| = |z_{3}|$. This condition enables the cancellation of the two leading terms in the determinant of the coefficient matrix for $C_i$ under open boundary conditions for $N \rightarrow \infty$, as is required for nontrivial solutions for $C_i$. Since the phases of $z_2,z_3$ in the GBZ are unspecified, the solutions form a continuum set of solutions on the $E$-plane, which are the bulk OBC bands~\cite{yokomizo2019nonbloch,yao2018edge,wang2024nonhermitian}. 

We can study the eigenvector topology more easily by mapping the bulk eigenvector to $M\in \mathbb{C}$ where $M$ is the ratio of the two bulk eigenvector elements $M(z) = a(z)/ b(z)$ which can be found to be~\cite{zhong2024polezero}:
\begin{equation}
    M(z,E)=\frac{d_x(z)-i d_y(z)}{E-d_0(z)-d_z(z)}=\frac{E-d_0(z)+d_z(z)}{d_x(z)+i d_y(z)}.
    \label{Eq:M}
\end{equation}
where $E$ satisfies Eq.~\eqref{Eq:charpoly}. 
For Hermitian systems with chiral or sublattice symmetry and where the model is in off-diagonal form, the typical topological invariant is the Berry-Zak phase~\cite{chiu2016classification,asboth2015ashort,zhong2024polezero}: 
\begin{equation}
    W_{\text{BZ,chiral}}^{(i)}=\oint_{\mathcal{C}_{\text{BZ}}^{(i)}} \frac{1}{2 \pi} \frac{d}{d z} \arg [M(z,E^{(i)})] d z
    \label{Eq:chiral}
\end{equation}
where the integration contour $\mathcal{C}_{\text{BZ}}^{(i)}$ is the Brillouin zone $|z| = 1$ (the unit circle on the $z$-plane) and $i=1,2$ is the band index. Although $E^{(i)}$ from the two bands are distinct and the $M$ values are different, $W_{\text{BZ,chiral}}^{(1)} = W_{\text{BZ,chiral}}^{(2)}$ due to mode orthogonality. For nearest-neighbor coupling models,
$W_{\text{BZ,chiral}}^{(i)} = 0$ or $\pm1$ corresponds to the absence or presence of a pair of topological edge states respectively~\cite{chiu2016classification,asboth2015ashort}. A form similar to Eq.~\eqref{Eq:chiral} can be given for non-Hermitian sublattice symmetric models by replacing the integration contour with $\mathcal{C}_{\text{GBZ}}^{(i)}$, the GBZ associated with either one of the bands~\cite{yokomizo2019nonbloch,yao2018edge,yang2020nonhermitian,zhong2024polezero}.

We can interpret $M$ on the Riemann sphere via the stereographic projection:
\begin{align}
X(M) = \frac{2  \Re(M)}{|M|^2 + 1}, Y(M) = \frac{2  \Im(M)}{|M|^2 + 1},Z(M) = \frac{|M|^2 - 1}{|M|^2 + 1}\label{Eq:MRiemann}
\end{align}
where $(X,Y,Z)$ are Cartesian coordinates of a point on the Riemann sphere and we will refer to the sphere as defined by Eq.~\eqref{Eq:MRiemann} as the $M$-Riemann sphere. $M(z,E^{(i)})$ in Eq.~\eqref{Eq:chiral} defines the image of the bulk bands on the $M$-plane, denoted as $M(\mathcal{C}_{\text{BZ}})$ which can be plotted on the $M$-Riemann sphere. If the image of the two bands on the $M$-Riemann sphere each winds around the north-south pole axis, then the system supports a pair of topological edge states. In Fig.~\ref{fig:couplings}(b) we show $M(\mathcal{C}_{\text{BZ}})$ on the $M$-Riemann sphere for a trivial SSH and topological SSH model respectively. We see that it does not wind around the north-south pole axis of the $M$-Riemann sphere for the trivial case but does for the topological case. We also plot the OBC spectra for a finite $N=32$ model and note that we have zero-energy edge states present only in the topological SSH, consistent with previous studies~\cite{chiu2016classification,asboth2015ashort}. The interpretation of the $M$-Riemann sphere is the same as two-band models on the Bloch sphere for Hermitian models~\cite{asboth2015ashort}, and a non-Hermitian generalization of the Bloch sphere for non-Hermitian systems~\cite{lieu208topological,zhong2024polezero}.

Eq.~\eqref{Eq:chiral} is not applicable for asymmetric models. In Fig.~\ref{fig:couplings}(c) we consider several cases of such asymmetric Hermitian Hamiltonians (the parameters are given in the End Matter). We see the presence of edge states even though the image of the two bands do not wind around the north-south pole axis. The observations here motivate us to develop a general topological invariant that accounts for the existence of edge states in these asymmetric Hamiltonians.

\textit{The Hermitian invariant -- } We describe a topological invariant applicable to all gapped Hermitian models of Eq.~\eqref{Eq:H}. We introduce the concept of a \textit{bulk eigenvector degeneracy}, which forms the basis of our invariant. From Eq.~\eqref{Eq:charpoly}, an OBC eigenstate at energy $E$ consists of a linear combination of the four $z$ solutions. Each $(z,E)$ pair can be mapped to $M$ using Eq.~\eqref{Eq:M}. At a given energy $E$, let us index the four $M_i$ solutions with the corresponding $z_i$ it was mapped from. A bulk eigenvector degeneracy is when $M_i =M_j \equiv M_{\text{deg}}$ but $z_i \neq z_j$. We show in the End Matter that $M_{\text{deg}}$ can be solved from:
\begin{equation}
    \begin{aligned}
& M_{\text{deg}}^2\left(t_{a a, 1} t_{b a,-1}-t_{a a,-1} t_{b a, 1}\right) +\left(t_{a b, 1} t_{b b,-1}-t_{a b,-1} t_{b b, 1}\right)+\\
& M_{\text{deg}}\left(t_{a b, 1} t_{b a,-1}-t_{a b,-1} t_{b a, 1}+t_{a a, 1} t_{b b,-1}-t_{a a,-1} t_{b b, 1}\right) =0.
\label{Eq:Mdeg}
\end{aligned}
\end{equation}
Thus, there are in general two $M_{\text{deg}}$ solutions for our model. 

We now generalize Eq.~\eqref{Eq:chiral} as
\begin{equation}
    W_{\text{BZ}}^{(i)}=\left( \sum_j 
\!\oint\limits_{\mathcal{C}_{\text{BZ}}^{(i)}} \!\frac{d}{d z} \arg \left[M(z,E^{(i)})-M_{\text{deg},j}\right] \frac{dz}{2\pi}  \right) \bmod 2.
    \label{Eq:restricted}
\end{equation}
Each integral can be interpreted as a winding number of the image of a single band, i.e. $M(\mathcal{C}^{(i)}_{\text{BZ}})$ around $M_{\text{deg},j}$ on the $M$-plane.
The sum over $j$ adds the contribution from each bulk eigenvector degeneracy point. After taking mod 2,  $W_{\text{BZ}}^{(i)}$ can be 0 or 1. The total winding number $W$ is defined as $W=W_{\text{BZ}}^{(1)}+W_{\text{BZ}}^{(2)}$ (without a mod 2). We claim that $W$ corresponds to the number of (possibly non-zero energy) edge states, and the model under consideration can only have 0, 1 or 2 edge states.

For Hermitian models that are sublattice-symmetric  and in off-diagonal form, $M_{\text{deg}}$ is 0 or $\infty$~\cite{zhong2024polezero} and Eq.~\eqref{Eq:restricted} reduces to Eq.~\eqref{Eq:chiral} with nearest neighbour couplings only. In asymmetric models, $M_{\text{deg}}$ can be arbitrary points $\in \mathbb{C}$ and different from 0 or $\infty$. Examples of asymmetric cases for Hermitian models are given in Fig.~\ref{fig:couplings}(c). The cases shown in Fig.~\ref{fig:couplings}(c) correspond to 0, 1 or 2 edge states. Calculating $W$ from Eq.~\eqref{Eq:restricted} amounts to counting how many $M(\mathcal{C}_{\text{BZ}})$ loop components can be continuously deformed to enclose an infinitesimal region containing exactly one $M_{\text{deg}}$ point on the $M$-Riemann sphere. These correspond to $W=0$, $1$, and $2$ in Fig.~\ref{fig:couplings}(c) respectively, which also correspond to the number of edge states in the OBC.

\begin{figure}[t]
\centering
\includegraphics[width=0.5\textwidth]{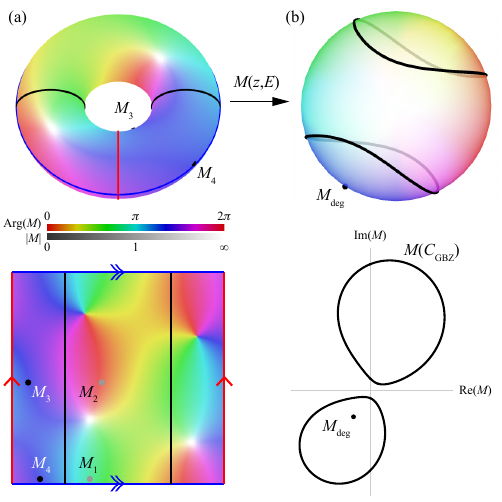}
\caption{General invariant as a winding number resulting from the mapping between Riemann surfaces. The Hamiltonian is the same as Fig. \ref{fig:couplings}(c), middle panel.
(a) Top panel: Band structure Riemann surface as a torus, colored by $M$ values where the color gives $\arg(M)$ and the brightness gives $|M|$. Black curves are the GBZ on the torus. Bottom panel: planar representation of the torus by cutting along the red and blue circles. The set of $M$ values for the only edge state in the model, where $M_3 = M_4$, have been marked on the plane and on the torus ($M_3$ and $M_4$ only).
(b) Top panel: The $M$-Riemann sphere with the $M_\text{deg} = M_3 = M_4$ (black dot) point and the $M(\mathcal{C}_{\text{GBZ}})$ (black curves). Bottom: Representation on the $M$ plane, showing the $M(\mathcal{C}_{\text{GBZ}})$ and the $M_\text{deg}$ point.}
\label{fig:proof}
\end{figure}
\textit{The general invariant -- } Eq.~\eqref{Eq:restricted} only works in the Hermitian case where the GBZ reduces to the BZ. In generic non-Hermitian systems, the GBZ may have a different topology, possibly leading to nontrivial braiding of the OBC bands~\cite{fu2024braiding,li2022topological} or disconnected components of the GBZ~\cite{wang2024nonhermitian}. We now present the following invariant that determines edge states of generic models in the form of Eq.~\eqref{Eq:H} regardless of the underlying GBZ eigenvalue topology. We define
\begin{equation}
    W_{j} \equiv \left(1 + \frac{1}{2 \pi i} \!\!\int\limits_{M(\mathcal{C_{\text{GBZ}}})} \!\! d M \frac{d}{d M} \ln \frac{M-M_{\text {deg},j}}{M-M_{\mathrm{branch}}}\right) \bmod 2
    \label{Eq:generalinvariant}
\end{equation}
where $M_{\mathrm{branch}}$ are the $M$-plane branch points (see End Matter for derivation). An $M_{\text{deg},j}$ point corresponds to an edge state if $W_{j} = 1$. The total number of edge states is $W = W_{1} + W_{2}$, and reduces to Eq.~\eqref{Eq:restricted} for Hermitian systems.

\textit{Sketch of the proof--} We now provide a sketch of the proof of Eqs.~\eqref{Eq:restricted} and \eqref{Eq:generalinvariant} by considering the analytical condition for edge states in the OBC for a finite chain with $N$ unit cells. For a nontrivial solution of the model in Eq.~\eqref{Eq:H} under OBC, the determinant of the coefficient matrix of $C_i$ must be zero, leading to the condition (see End Matter for more details):
\begin{align}
    & \left|\begin{array}{ll}
M_1 & M_2 \\
1 & 1
\end{array}\right|\left|\begin{array}{ll}
M_3 & M_4 \\
1 & 1
\end{array}\right| - \left|\begin{array}{ll}
M_1 & M_3 \\
1 & 1
\end{array}\right|\left|\begin{array}{cc}
M_2 & M_4 \\
1 & 1
\end{array}\right| \left(\frac{z_2}{z_3}\right)^N \nonumber\\
    &+ \cdots=0. \label{Eq:analyticalOBC}
\end{align}
If Eq.~\eqref{Eq:analyticalOBC} holds in the thermodynamic limit $N \rightarrow \infty$, the leading terms shown here must vanish. One way to do so is to have $|z_2|=|z_3|$ which is the GBZ condition described earlier (the bulk states). The other way is to have $M_1=M_2$ or $M_3=M_4$ so that the leading term is zero. These lead to states with discrete and isolated energy, and this is therefore the analytical condition for edge states. In the End Matter, we can show that for Hermitian models, $M_1=M_2$ ($M_3=M_4$) is a left-localized (right-localized) edge state. 

We note that $M_1=M_2$ or $M_3=M_4$ are two possible but not all cases of $M_{\text{deg}}$, as one can also have e.g. $M_1=M_3$. Eqs.~\eqref{Eq:restricted} and \eqref{Eq:generalinvariant} can be interpreted as an index tracker. When the winding number is nontrivial, the $M_{\text{deg}}$ points have the correct indices $i$ in $M_i$ that lead to edge states. The indices of the $M_{\text{deg}}$ points are related to the GBZ. Since the GBZ is defined by $|z_2| = |z_3|$, on the band structure Riemann surface (Fig~\ref{fig:proof}(a)), $z_1$ and $z_2$ will always be on the opposite side of the GBZ as compared with $z_3$ and $z_4$. Consider now the indices associated with the $M_{\text{deg}}$ point. As an example, suppose with certain parameters of the Hamiltonian this $M_{\text{deg}}$ point corresponds to $M_1 = M_3$. If we vary the parameters of the Hamiltonian, this $M_{\text{deg}}$ point may cross $M(\mathcal{C}_{\text{GBZ}})$, the image of the GBZ on the $M$-sphere. When the crossing occurs, however, the labeling of the $z$ values must change between $z_2$ and $z_3$, and therefore the indices associated with the $M_{\text{deg}}$ point must change such that it now corresponds to $M_1 = M_2$. The crossing therefore indicates the emergence of an edge state from the bulk state as defined by the GBZ (see SM Sec.~\ref{supp:visualM} for visualizations of this).

The proof of Eq.~\eqref{Eq:generalinvariant} formalizes the above argument by defining the \textit{inside of the GBZ}, consisting of states where the $z$'s are two roots of Eq.~\eqref{Eq:charpoly} with the smallest amplitudes (i.e. $z_1$ or $z_2$). The inside of the GBZ is considered as a region on the Riemann surface of the band structure, which is then analytically mapped to the $M$-Riemann sphere through Eq.~\eqref{Eq:M} (Fig. \ref{fig:proof}). The $M_{\text{deg}}$ point on the sphere will be covered twice by the GBZ inside if $M_1 = M_2$, zero times if $M_3 = M_4$, and exactly once for all other cases. By defining a winding integral on the band structure Riemann surface and mapping it to the $M$-Riemann sphere, the number of times the GBZ inside covers a region can be expressed as an $M$ winding number. The argument above directly leads to Eq.~\eqref{Eq:restricted}, applicable for Hermitian models. For non-Hermitian models, the $M(\mathcal{C}_{\text{GBZ}})$ loops may merge or intersect with each other, and it becomes necessary to take into account the contour direction used for integration. This is achieved by modifying the raw winding number and subtracting the winding number of $M_{\text{branch}}$, whose preimages are degenerate and therefore always on the same side of the GBZ. Complete proofs for Eqs.~\eqref{Eq:generalinvariant} and \eqref{Eq:restricted} can be found in SM Sec.~\ref{supp:generalinvariant} and \ref{supp:restrictedinvariant}, respectively.

\begin{figure}[t]
\centering
\includegraphics[width=0.45\textwidth]{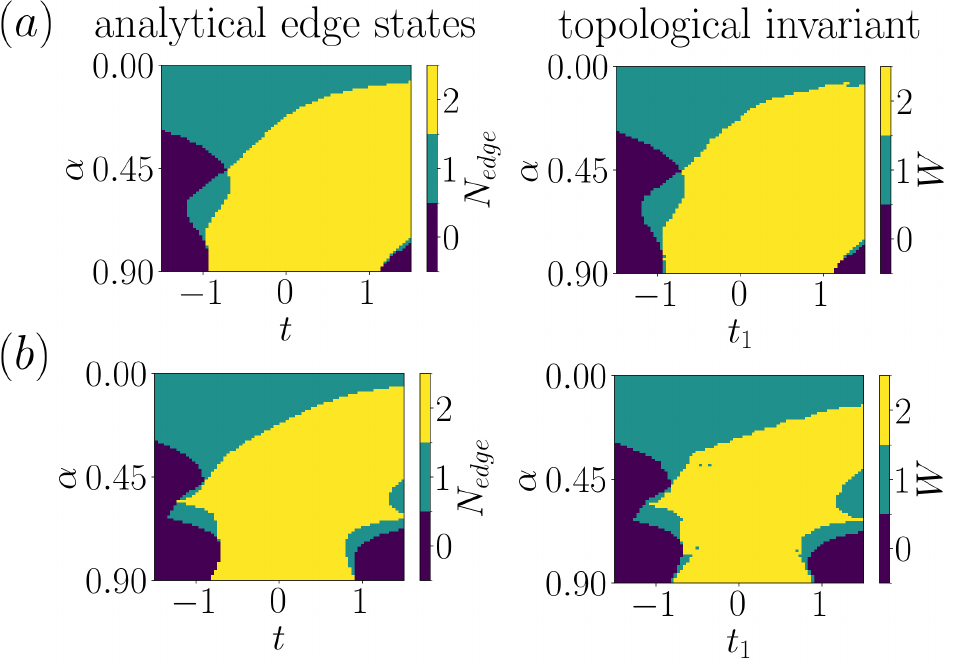}
\caption{(a) Analytical edge state and topological invariant phase diagram calculated from Eq.~\eqref{Eq:restricted} for finite chain with $N=32$ for 6400 models for a linear combination of a Hermitian random model and a Hermitian SSH model, where $\alpha$ is a weighting parameter. (b) Same as (a) but using Eq.~\eqref{Eq:generalinvariant} for topological invariant and using a linear combination of a non-Hermitian random model and a Hermitian SSH model, where $\alpha$ is still a weighting parameter. Full parameters are in the End Matter.}
\label{fig:phasediagram}
\end{figure}

\textit{Numerical results -- } We numerically demonstrate both the Hermitian invariant Eq.~\eqref{Eq:restricted} and the general invariant in Eq.~\eqref{Eq:generalinvariant} by comparing the topological invariant to the analytical criteria for edge states from Eq.~\eqref{Eq:analyticalOBC} in Fig.~\ref{fig:phasediagram}. Eq.~\eqref{Eq:analyticalOBC} predicts OBC edge states at energies where $|\frac{M_1}{M_2} - 1| = 0$ and $|\frac{M_3}{M_4} - 1| = 0$, where $M_1$ to $M_4$ is solved from from Eq.~\eqref{Eq:charpoly} and Eq.~\eqref{Eq:M}. In numerical solvers, these values do not vanish exactly so we classify a mode as an edge state if and only if $|\frac{M_1}{M_2} - 1| < \epsilon$ or $|\frac{M_3}{M_4} - 1| < \epsilon$, for some small tolerance $\epsilon$ (here, we use $\epsilon = 0.001$). We use a randomly chosen Hermitian and non-Hermitian model for Fig.~\ref{fig:phasediagram}(a) and Fig.~\ref{fig:phasediagram}(b) respectively called $H_{\text{h}}$ and $H_{\text{nh}}$ (full parameters given in the End Matter). In Fig.~\ref{fig:phasediagram}(a), we define $H(z) = (1-\alpha ) H_{\text{h}} + \alpha  H_{\text{ssh}}$, and we vary $\alpha $ from 0 to 0.90 and $t_1$ in the SSH Hamiltonian from -1.5 to 1.5 over 80 steps, resulting in 6400 models. We do a similar construction, except with the use of $H_{\text{nh}}$ for Fig.~\ref{fig:phasediagram}(b). The topological invariants (on the left in Fig.~\ref{fig:phasediagram}) are calculated by fitting a curve to $M(\mathcal{C}_{\text{GBZ}})$ and counting $M_{\text{deg},j}$ (and $M_{\text{branch}}$ for the general invariant) inside the $M(\mathcal{C}_{\text{GBZ}}$) curves as dictated by Eq.~\eqref{Eq:restricted} and Eq.~\eqref{Eq:generalinvariant}. For the analytical edge state phase diagram (right panels in Fig.~\ref{fig:phasediagram}), we used the $M_{\text{deg}}$ from Eq.~\eqref{Eq:Mdeg} and Eq.~\eqref{Eq:charpoly} and~\eqref{Eq:M} to find the $M_i$ indices it corresponded to. We then applied the $|\frac{M_1}{M_2} - 1| < \epsilon$, $|\frac{M_3}{M_4} - 1| < \epsilon$ criteria to find the number of edge states $N_{\text{edge}}$. In Fig.~\ref{fig:phasediagram}, we see that $N_{\text{edge}}$ and $W$ from the invariants match very well, with only slight differences in the boundaries due to finite-size effects or the choice of $\epsilon$. The analytical edge state condition in Eq.~\eqref{Eq:analyticalOBC} does not contain the winding information in Eq.~\eqref{Eq:restricted} and Eq.~\eqref{Eq:generalinvariant} and Fig.~\ref{fig:phasediagram} therefore verifies the validity of Eq.~\eqref{Eq:restricted} and Eq.~\eqref{Eq:generalinvariant}. More details on these numerical tests can be found in SM Sec.~\ref{supp:phasediagram}.

\textit{Discussion -- } Unlike many symmetry-protected topological invariants~\cite{chiu2016classification,yokomizo2019nonbloch,yao2018edge}, both Eq.~\eqref{Eq:restricted} and Eq.~\eqref{Eq:generalinvariant} remain unchanged under $z$-independent unitary and similarity transformations of the Hamiltonian. Such transformations correspond to Möbius transformations on the $M$-Riemann sphere~\cite{zhong2024polezero}, which preserves cross-ratios~\cite{bruna2013complex} and leaves the invariant unchanged. This invariance aligns with the physical intuition that edge states should be unaffected by such transformations as both the bulk spectral properties and the OBC are unaffected. 

The topological invariants we provide in this work are applicable to systems that are traditionally defined as a topological or trivial phase in symmetry-protected classifications~\cite{chiu2016classification}. Within the context of our models of Eq.~\eqref{Eq:H}, they are a broader invariant that applies to non-zero energy edge states and the emergence of which does not coincide with the gap closing condition (see SM Sec.~\ref{supp:nosymmetry}). Our work is related but distinct from several recent works defining a topological invariant without symmetry constraints using Green's function poles~\cite{rhim2018unified,tamura2021generalization,muller2021universal,piasotski2022universal, pletyukhov2020surface,pletyukhov2020topological} which requires a Green's function of a truncated system. In contrast, our invariant is calculated purely from bulk quantities. Additionally, our formalism gives a measure of the perturbation strength needed to form edge states as (unlike Green's function poles) bulk eigenvector degeneracy points exist even when not corresponding to edge states (see SM Sec.~\ref{supp:robustness}). The bulk eigenvector degeneracy points cannot be created or destroyed for general cases of Eq.~\eqref{Eq:H}, which gives the edge states associated with the bulk eigenvector degeneracies a topological origin under our formalism in Eq.~\eqref{Eq:restricted} and Eq.~\eqref{Eq:generalinvariant}.

In conclusion, we have derived new topological invariants that apply to asymmetric and symmetry-protected nearest-neighbor two-band models with arbitrary complex couplings. Our work demonstrates that the full complex band structure~\cite{rhim2018unified,pletyukhov2020surface,pletyukhov2020topological,alase2017generalization,wang2024onedimensional,wang2024nonhermitian,kunst2019nonhermitian,wielian2025transfer,koekenbier2024transfer,ghaemi2023transport,guarie2011single,essin2011bulk,mong2011edge,fulga2012scattering,montag2024topological,yang2020nonhermitian,hu2023greens,lee2019anatomy,prodan2006analytic,kohn1959analytic,reuter2017unified,hu2024geometric,hatsugai1993edge} is fundamental not only for non-Hermitian systems but also for Hermitian ones. The analytical edge state condition we use here can be extended to longer-range and continuous models (see SM Sec.~\ref{supp:secondnearest}). Thus, our understanding of the topological origins of edge states in asymmetric models may generalize to other Hermitian and non-Hermitian models~\cite{long2024non,verresen2018topology,longhi2018nonhermitian,stjean2017lasing,kitaev2001unpaired,stern2013topological,chiu2016classification,lu2014topological,xiao2014surface,tang2021topology,ni2022topological}.

J.Z. thanks Prof. Zhong Wang for helpful discussions at an early stage of this work. This work is supported by a Simons Investigator in Physics grant from the Simons Foundation (Grant No. 827065), and by a MURI grant from the U. S. Air Force Office of Scientific Research (Grant No. FA9550-22-1-0339).  The work of A.N.P. has been supported by research grants
from the Center for New Scientists and from the Center for
Scientific Excellence at the Weizmann Institute of Science, by the Quantum Science and Technology Program of the
Israel Council for Higher Education and by the Minerva Foundation. J.Z. was funded in part by the Fulbright Future Scholarship.


%

\setcounter{figure}{0}
\clearpage
\newpage
\onecolumngrid
\section*{End Matter}
\twocolumngrid
\textbf{Model parameters: } We write Eq.~\eqref{Eq:H} in the form $H(z) = h_- z^{-1} + h_0 + h_+ z$ where
\begin{align}
\begin{gathered}
    h_{-}=\left[\begin{array}{cc}
t_{a a,-1} & t_{a b,-1} \\
t_{b a,-1} & t_{b b,-1}
\end{array}\right], h_0=\left[\begin{array}{ll}
t_{a a, 0} & t_{a b, 0} \\
t_{b a, 0} & t_{b b, 0}
\end{array}\right], \\
h_{+}=\left[\begin{array}{cc}
t_{a a, 1} & t_{a b, 1} \\
t_{b a, 1} & t_{b b, 1}
\end{array}\right] .
\end{gathered}
\end{align}

\begin{table*}
\centering
\begin{tabular}{|c|c|c|c|}
\hline
Model& \( h_- \) & \( h_0 \) & \( h_+ \) \\
\hline
$H_{\text{h}}$& 
\(\begin{bmatrix}
0.84 - 0.14i & 0.02 - 0.69i \\
0.02 - 0.69i & -0.82 - 0.60i
\end{bmatrix}\) & 
\(\begin{bmatrix}
-0.66 + 0.00i & -0.96 - 0.80i \\
-0.96 + 0.80i & -0.58 + 0.00i
\end{bmatrix}\) & 
\(\begin{bmatrix}
0.84 + 0.14i & 0.02 + 0.69i \\
0.02 + 0.69i & -0.82 + 0.60i
\end{bmatrix}\) \\
\hline
$H_{\text{nh}}$ & 
\(\begin{bmatrix}
0.74 - 0.13i & -0.79 - 0.84i \\
-0.43 - 0.08i & 0.01 + 0.60i
\end{bmatrix}\) & 
\(\begin{bmatrix}
-0.18 + 0.32i & -0.44 - 0.84i \\
-0.33 + 0.70i & 0.21 - 0.74i
\end{bmatrix}\) & 
\(\begin{bmatrix}
0.51 + 0.40i & -0.86 + 0.40i \\
-0.44 + 0.38i & -0.86 - 0.03i
\end{bmatrix}\) \\
\hline
\( H_{\text{ssh}} \) & 
\(\begin{bmatrix}
0 & t_2 \\
0 & 0
\end{bmatrix}\) & 
\(\begin{bmatrix}
0 & t_1 \\
t_1 & 0
\end{bmatrix}\) & 
\(\begin{bmatrix}
0 & 0 \\
t_2 & 0
\end{bmatrix}\) \\
\hline
\end{tabular}
\caption{Model parameters used in Figs.~\ref{fig:couplings} and~\ref{fig:phasediagram}.}
\label{table:parametersFig1}
\end{table*}

The models used in Figs.~\ref{fig:couplings} and~\ref{fig:phasediagram} are described in Table ~\ref{table:parametersFig1}. In the two panels of Fig.~\ref{fig:couplings}(b) we use $H=H_{\text{ssh}}$ with $t_1=1.5$ and $0.5$, respectively, and $t_2=1$ for both panels. In the three panels of Fig.~\ref{fig:couplings}(c) we use
$H=(1-\alpha)H_{\text{h}}+\alpha  H_{\text{ssh}}$ where $t_1 = 1.5, t_2 = 1, \alpha  = 0.85$ for the case with no edge states, $\alpha=0$ for the case with one edge state and $t_1 = 0.5, t_2 = 1, \alpha  = 0.5$ for the case with two edge states. The parameters for these cases are within the parameter ranges of the phase diagram in Fig.~\ref{fig:phasediagram}(a). In Fig.~\ref{fig:phasediagram}(a), we use $H=(1-\alpha )H_{\text{h}}+\alpha H_{\text{ssh}}$, where $t_2=1$ for the SSH parameter, and $\alpha$ and $t_1$ vary. In Fig.~\ref{fig:phasediagram}(b) we use $H=(1-\alpha )H_{\text{nh}}+\alpha  H_{\text{ssh}}$, where $t_2=1$, and $\alpha$ and $t_1$ vary.\\

\textbf{$M_{\text{deg}}$ and $M_{\text{branch}}$ derivation: } The degeneracy point $M_{\text{deg}}$ (and the $E_{\text{deg}}$ value at which $M_{\text{deg}}$ occurs) can be found by simultaneously solving the four equations $M_{\text{eqn1}}(z_a,E)=M_{\text{eqn2}}(z_a,E)=M_{\text{eqn1}}(z_b,E)=M_{\text{eqn2}}(z_b,E)=M$ for the four unknowns $M\equiv M_{\text{deg}}, E \equiv E_{\text{deg}}, z_a,z_b$ where $M_{\text{eqn1}}(z,E)$ and  $M_{\text{eqn2}}(z,E)$ are the two forms of Eq.~\eqref{Eq:M} and $z_a \neq z_b$. The solution for $M$ gives Eq.~\eqref{Eq:Mdeg} (more details can be found in SM Sec.~\ref{supp:MdegEdeg}). Using the two forms of Eq.~\eqref{Eq:M}, we can also eliminate $E$ and arrive at a polynomial equation $g(M,z)=0$ which is a degree 2 polynomial in $M$ and a degree 2 polynomial in $z$. $M_{\text{branch}}$ used in Eq.~\eqref{Eq:generalinvariant} can be found by calculating the roots of the discriminant of $g(M, z)=0$ with respect to $z$ (more details can be found in SM Sec.~\ref{supp:riemann}). \\

\textbf{Analytical OBC: }We explain how to derive Eq.~\eqref{Eq:analyticalOBC}. From the generalized Bloch ansatz, a general mode solution for Eq.~\eqref{Eq:H} in the presence of a boundary at energy $E$ is a superposition of all the $p=4$ states:
\begin{equation}
\binom{\psi_{na}}{\psi_{nb}}=C_1\binom{a_1}{b_1} z_1^n+C_2\binom{a_2}{b_2} z_2^n+C_3\binom{a_3}{b_3} z_3^n+C_4\binom{a_4}{b_4} z_4^n
\label{Eq:obcztransform}
\end{equation}
We now impose open-boundary conditions (OBC) by requiring that $\psi_{na}=\psi_{nb}=0$ for $n=0$ and $n=N$ which gives:
\begin{equation}
    \left(\begin{array}{cccc}
a_1 & a_2 & a_3 & a_4 \\
b_1 & b_2 & b_3 & b_4 \\
a_1 z_1^N & a_2 z_2^N & a_3 z_3^N & a_4 z_4^N \\
b_1 z_1^N & b_2 z_2^N & b_3 z_3^N & b_4 z_4^N
\end{array}\right)\left(\begin{array}{c}
C_1 \\
C_2 \\
C_3 \\
C_4
\end{array}\right)=\left(\begin{array}{l}
0 \\
0 \\
0 \\
0
\end{array}\right).
\label{Eq:NNcoefficient}
\end{equation}
For Eq.~\eqref{Eq:obcztransform} to have nonzero solutions, the determinant of the coefficient matrix must be zero~\cite{alase2017generalization}. We expand the determinant and order the terms by the magnitudes of $z$:
\begin{equation}
    \left|\begin{array}{cc}a_1 & a_2 \\ b_1 & b_2\end{array}\right|\left|\begin{array}{cc}a_3 & a_4 \\ b_3 & b_4\end{array}\right| z_3^{N} z_4^N-\left|\begin{array}{cc}a_1 & a_3 \\ b_1 & b_3\end{array}\right|\left|\begin{array}{cc}a_2 & a_4 \\ b_2 & b_4\end{array}\right| z_2^N z_4^N+\cdots=0
\end{equation}
and using $M_i=a_i /b_i$, we get Eq.~\eqref{Eq:analyticalOBC}. For Eq.~\eqref{Eq:analyticalOBC} to hold, the leading terms should cancel in the $N\rightarrow\infty$ limit. The first way is to require $|z_2|/|z_3| \rightarrow 1$ and results in the condition for the GBZ (the bulk states). The second way is to require the first term to cancel and gives $(M_1-M_2)(M_3-M_4) \rightarrow 0$ which leads to either $M_1=M_2$ or $M_3=M_4$ (the edge states). 

It is possible to analytically solve for $C_1$ to $C_4$~\cite{hou2023topological,yokomizo2019nonbloch} which will give us the analytical mode profile using Eq.~\eqref{Eq:obcztransform} with $M_i=a_i/b_i$ which is $\binom{\psi_{n a}}{\psi_{n b}}=\sum_{i=1}^p C_i z_i^n\binom{M_i}{1}$ where $n$ labels the unit-cell index. Here we solve it using $C_i=(-1)^{1+i}  g  \operatorname{det}\left(A_{1 i}\right)$ where $A_{1 i}$ is the minor matrix obtained by deleting the first row and $i$-th column of the coefficient matrix and $g$ is an arbitrary scaling constant which gives Eq.~\eqref{Eq:C1} to~\eqref{Eq:C4}.

\begin{widetext}
\begin{align}
C_1 z_1^n &= g \left( M_4 \left|\begin{array}{ll} M_3 & M_2 \\ 1 & 1 \end{array}\right| z_1^n z_2^N z_3^N - M_3 \left|\begin{array}{ll} M_4 & M_2 \\ 1 & 1 \end{array}\right| z_1^n z_2^N z_4^N + M_2 \left|\begin{array}{ll} M_4 & M_3 \\ 1 & 1 \end{array}\right| z_1^n z_3^N z_4^N \right) \label{Eq:C1}\\
C_2 z_2^n &= g \left( M_4 \left|\begin{array}{ll} M_1 & M_3 \\ 1 & 1 \end{array}\right| z_1^N z_2^n z_3^N + M_3 \left|\begin{array}{ll} M_4 & M_1 \\ 1 & 1 \end{array}\right| z_1^N z_2^n z_4^N - M_1 \left|\begin{array}{ll} M_4 & M_3 \\ 1 & 1 \end{array}\right| z_2^n z_3^N z_4^N \right) \\
C_3 z_3^n &= g \left( M_4 \left|\begin{array}{ll} M_2 & M_1 \\ 1 & 1 \end{array}\right| z_1^N z_2^N z_3^n - M_2 \left|\begin{array}{ll} M_4 & M_1 \\ 1 & 1 \end{array}\right| z_1^N z_3^n z_4^N + M_1 \left|\begin{array}{ll} M_4 & M_2 \\ 1 & 1 \end{array}\right| z_2^N z_3^n z_4^N \right) \\
C_4 z_4^n &= g \left( M_3 \left|\begin{array}{ll} M_1 & M_2 \\ 1 & 1 \end{array}\right| z_1^N z_2^N z_4^n + M_2 \left|\begin{array}{ll} M_3 & M_1 \\ 1 & 1 \end{array}\right| z_1^N z_3^N z_4^n - M_1 \left|\begin{array}{ll} M_3 & M_2 \\ 1 & 1 \end{array}\right| z_2^N z_3^N z_4^n \right)\label{Eq:C4}
\end{align}
\end{widetext}

Let $n=kN$ for any $0<k<1$ be in the chain away from the ends and $N \rightarrow \infty$. When $\binom{\psi_{n a}}{\psi_{n b}}$ is dominated by one $C_i z_i^n\binom{M_i}{1}$ term, $|z_i|$ contains the localization information of the eigenstate since $\binom{\psi_{n a}}{\psi_{n b}} \approx C_i z_i^n\binom{M_i}{1}=C_ie^{i \operatorname{Re}\left(k_i\right) n} e^{-\operatorname{Im}\left(k_i\right) n}\binom{M_i}{1}$ and $\text{Im}(k_i)=-\log(|z_i|)$ is the amplification or decay factor of the eigenstate. Assume that if there are two edge states, they are not at degenerate energies or close in energy (this excludes many symmetric model edge states like SSH edge states). For $M_1=M_2$ in this limit, the leading term gives $\binom{\psi_{n a}}{\psi_{n b}} \approx C_2 z_2^n\binom{M_2}{1}$ and the mode has a localization length~\cite{asboth2015ashort} and direction $\xi \approx -1/\log(|z_2|)$. For $M_3=M_4$, $\binom{\psi_{n a}}{\psi_{n b}} \approx C_3 z_3^n\binom{M_3}{1}$ and $\xi \approx -1/\log(|z_3|)$. For $|z_2|=|z_3|$, $\binom{\psi_{n a}}{\psi_{n b}} \approx \sum_{i=2}^3 C_i z_i^n\binom{M_i}{1}$ and $\xi \approx -1/\log(|z_2|)= -1/\log(|z_3|)$. In Hermitian models, the Hermiticity ensures that if $z_i$ is a solution to Eq.~\eqref{Eq:charpoly} at energy $E$, then $(z_i^*)^{-1}$ is also a solution. Thus, $|z_1 z_4|=1$ and $|z_2 z_3|=1$. If $E$ is not a bulk state, then $|z_2|\neq1$ and $|z_3|\neq1$ and the edge states must have $|z_2|<1$ and $|z_3|>1$. Thus, we can conclude that $M_1=M_2$ is a left-localized edge state and $M_3=M_4$ is a right-localized edge state if they are at non-degenerate energies. For degenerate or close to degenerate edge states, the determinants in Eq.~\eqref{Eq:C1} to~\eqref{Eq:C4} can be small and the leading term analysis above can change such that edge states are hybridized modes. For non-Hermitian models, the edge and bulk states may change localization due to imaginary gauge transforms~\cite{longhi2018nonhermitian}, however, the relative localization strengths of edge states compared to the bulk states can be quantified by comparing $|z_2|$ (when $M_1=M_2$) and $|z_3|$ (when $M_3=M_4$) for edge states against the $|z_2|=|z_3|$ values of bulk states (noting that these $z_2,z_3$ values depends on $E$ of the mode as given by Eq.~\eqref{Eq:charpoly}). See SM Sec.~\ref{supp:modeprofiles} for mode profile examples.\\

\textbf{Longer-range model discussion: } For tight-binding models with longer coupling ranges, the bulk eigenvector degeneracy points discussed in the main text take modified forms, but can be expressed as roots of algebraic equations involving the coupling coefficients. By B\'ezout's theorem, there are a finite and fixed number of such points on the band structure when the coupling parameter changes. The case of next-nearest neighbor coupling is discussed in SM Sec.~\ref{supp:secondnearest}, where the transition of bulk eigenvector degeneracy points into and away from the edge-state condition involves crossing the GBZ curve, analogous to the nearest-neighbor case in the main text. This suggests that edge states in longer-range tight-binding asymmetric models may also have topological origins arising from bulk eigenvector degeneracy points and the GBZ, and a detailed investigation is left for future work.

\onecolumngrid 

\newpage
\setcounter{figure}{0}
\renewcommand{\thefigure}{S\arabic{figure}}
\setcounter{equation}{0}
\renewcommand{\theequation}{S\arabic{equation}}

\section*{Supplementary Material}

\section{Asymmetric models}
\label{supp:nosymmetry}
We give an example of an asymmetric model, as defined in the main text. Randomly generated models obeying Eq.~\eqref{Eq:H} such as $H_{\text{h}}$ do not obey symmetry constraints in their energy bands~\cite{mccann2023catalog} and are thus asymmetric models. In Fig.~\ref{fig:brokensym}(a) and (b), we plot $H=(1-\alpha) H_{\mathrm{h}}+\alpha H_{\mathrm{ssh}}$ with SSH parameters $t_1=0.5, t_2=1$ and $t_1=1.5, t_2=1$ respectively for $N=100$ unit cell models. Both figures are one slice of the phase diagram in Fig.~\ref{fig:phasediagram}(a). We color the eigenvalues by the inverse participation ratio (IPR) of the corresponding eigenvector~\cite{ganeshan2015nearest} where  $\mathrm{IPR}=\sum_m\left|\psi_m\right|^4$ and $\psi_m$ is a component of the $2N$ component real-space eigenvector of the model under OBC, normalized by $\sum_m\left|\psi_m\right|^2 = 1$. At $\alpha=1$, both models reduce to the $H_{\mathrm{ssh}}$ model, which has chiral symmetry. For $\alpha=1$, edge states under OBC are at zero energy and are found at exactly the middle of the gap between the bulk OBC bands in the $N\rightarrow \infty$ limit. As $\alpha$ decreases from 1, the model increasingly resembles $H_{\mathrm{h}}$, effectively introducing symmetry-breaking perturbations to $H_{\mathrm{ssh}}$. 

\begin{figure}[H]
    \centering
    \includegraphics[width=0.6\linewidth]{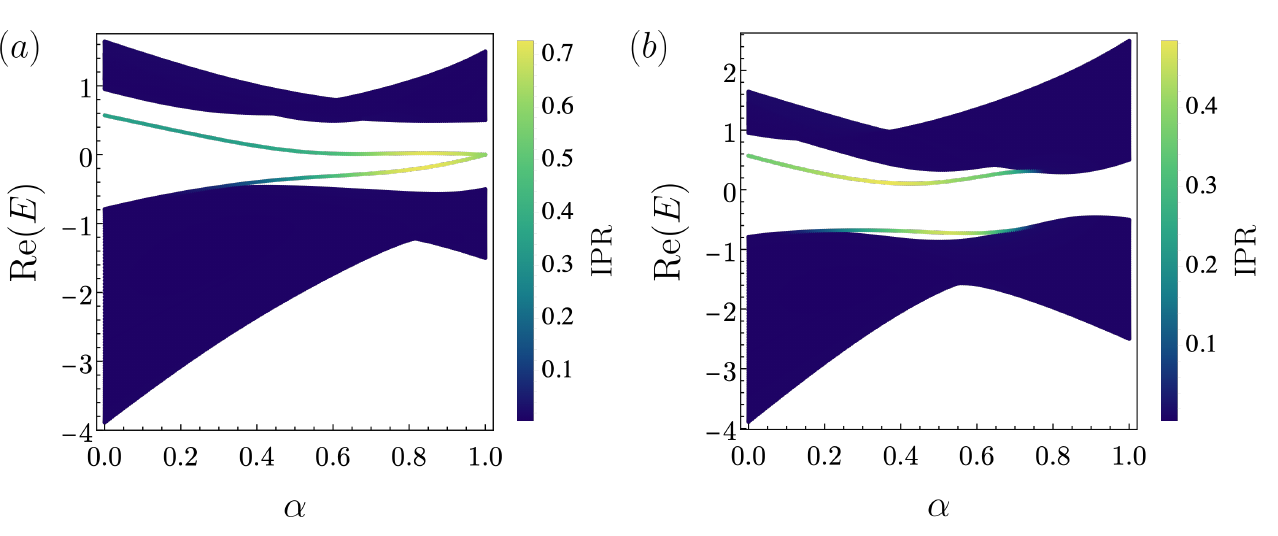}
    \caption{(a) Real-space OBC eigenvalues of $H=(1-\alpha) H_{\mathrm{h}}+\alpha H_{\mathrm{ssh}}$ with SSH parameters $t_1=0.5, t_2=1$ for $N=100$ unit cells where the eigenvalue is colored by the IPR of the corresponding eigenvector. (b) same as (a) but for SSH parameters $t_1=1.5, t_2=1$.}
    \label{fig:brokensym}
\end{figure}

\section{Eigenvalue and eigenvector Riemann surface}
\label{supp:riemann}

\subsection{Generalized Bloch theorem}
\label{supp:generalblochderivation}
Here we derive the generalized Bloch matrix in Eq.~\eqref{Eq:H} from first principles and demonstrate where the generalized Bloch theorem is applied (and why it applies for arbitrary boundary conditions) as well as where the characteristic polynomial in Eq.~\eqref{Eq:charpoly} comes from. The starting point is the real-space Hamiltonian for a tight-binding model described in Fig.~\ref{fig:couplings}(a). The discrete real-space equations of motion in the bulk give a set of coupled equations~\cite{longhi2019probing,wang2024nonhermitian,guo2021exact}:
\begin{align}
& t_{a a, 1} \psi_{(n+1)a} + t_{a a, 0} \psi_{na} + t_{a a,-1} \psi_{(n-1)a} + t_{a b, 1} \psi_{(n+1)b} + t_{a b, 0} \psi_{nb} + t_{a b,-1} \psi_{(n-1)b} = E \psi_{na} \\
& t_{b a, 1} \psi_{(n+1)a} + t_{b a, 0} \psi_{na} + t_{b a,-1} \psi_{(n-1)a} + t_{b b, 1} \psi_{(n+1)b} + t_{b b, 0} \psi_{nb} + t_{b b,-1} \psi_{(n-1)b} = E \psi_{nb}.
\end{align}
These equations apply to sites that are in the bulk~\cite{wang2024nonhermitian,guo2021exact,alase2017generalization}. For OBC, the boundary sites give the equations:
\begin{align}
&t_{a a, 0} \psi_{1a}+t_{a b, 0} \psi_{1b}+t_{a a, 1} \psi_{2a}+t_{a b, 1} \psi_{2b} =E \psi_{1a} \\
&t_{b a, 0} \psi_{1a}+t_{b b, 0} \psi_{1b}+t_{b a, 1} \psi_{2a}+t_{b b, 1} \psi_{2b} =E \psi_{1b} \\
&t_{a a,-1} \psi_{(N-1)a}+t_{a b,-1} \psi_{(N-1)b}+t_{a a, 0} \psi_{Na}+t_{a b, 0} \psi_{Nb} =E \psi_{Na} \\
&t_{b a,-1} \psi_{(N-1)a}+t_{b b,-1} \psi_{(N-1)b}+t_{b a, 0} \psi_{Na}+t_{b b, 0} \psi_{Nb} =E \psi_{Nb}
\end{align}

The bulk equations can be applied to the boundary sites by extending the lattice with $\psi_{0a}, \psi_{0b}, \psi_{(N+1)a}, \psi_{(N+1)b}$. Then, for the OBC, we set $\psi_{0a}= \psi_{0b}=\psi_{(N+1)a}=\psi_{(N+1)b}=0$~\cite{wang2024nonhermitian,guo2021exact}. The technique of applying bulk equations at boundary sites is also applicable for generalized boundary conditions including boundary impurities~\cite{alase2017generalization,li2023scale}, periodic boundaries (PBC)~\cite{guo2021exact}, OBC-PBC interpolations~\cite{guo2021exact,wang2024nonhermitian}, and semi-infinite boundaries~\cite{wang2024nonhermitian,okuma2020topological}. In these cases, the extended sites generally have non-zero amplitudes. By requiring the bulk equations (linear recurrence relations) to remain valid at all existing sites, the boundary conditions manifest themselves as constraints on the auxiliary amplitudes of the extended lattice. This is why the generalized Bloch ansatz applies even though translational symmetry may be broken by boundaries (unlike the typical Bloch ansatz $\left(\psi_{n a}, \psi_{n b}\right)^T=e^{ikn}\left(\psi_{0 a}, \psi_{0 b}\right)^T$, where a real $k$ corresponds to translational symmetry across the whole chain and is assumed to generally apply only for PBC).

The approach to solving linear recurrence relations is to take an ansatz $(\psi_{na}, \psi_{nb})^T = z^n (\psi_{0a},\psi_{0b})^T$, where $z\in \mathbb{C}$. Substituting this into the recurrence relations leads to the characteristic polynomial in Eq.~\eqref{Eq:charpoly} and the generalized Bloch matrix Eq.~\eqref{Eq:H}. There will be $p$ solutions of $z$ at some $E$ from the characteristic polynomial. Therefore, a general solution to the order $p$ linear recurrence relations is a sum of the $p$ solutions:
\begin{equation}
    \binom{\psi_{na}}{\psi_{nb}}=\sum_{i=1}^p C_i z_i^n\binom{M_i}{1}
\end{equation}
This is essentially the $z$-transform (or generalized Fourier transform or discrete Laplace transform)~\cite{bruna2013complex}. The boundary conditions determine $C_i$, which are analogous to Fourier amplitudes~\cite{wang2024nonhermitian} but complex $k$ Fourier amplitudes are allowed.

\subsection{Riemann surfaces and branch points}

\begin{figure}[H]
\centering
\includegraphics[width=\textwidth]{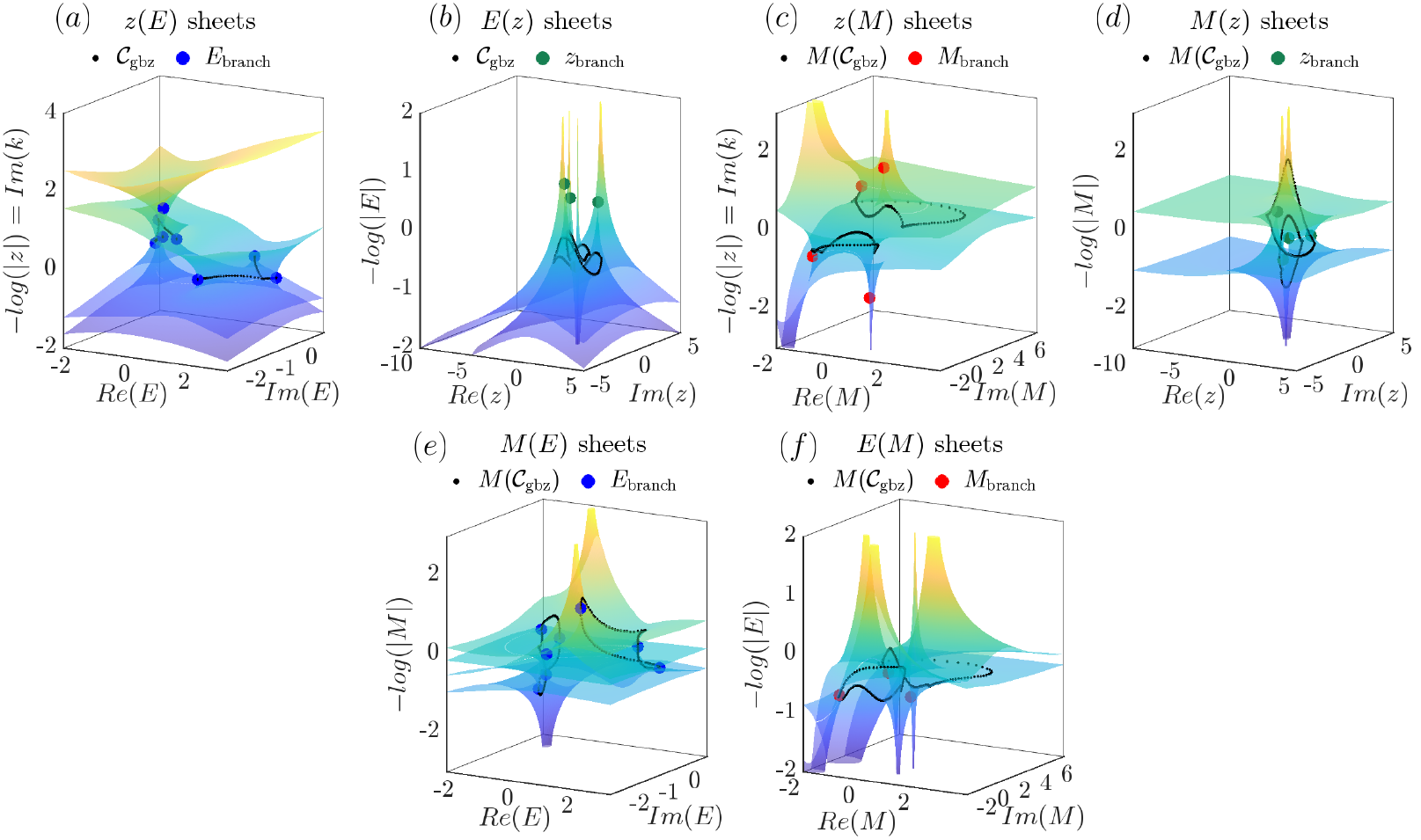}
\caption{In general, the nearest-neighbor (between unit cells) two-band tight-binding model in Eq.~\eqref{Eq:H} will have (a) four $z(E)$ sheets with eight $E$ branch points (b) two $E(z)$ sheets with four $z$ branch points (c) two $z(M)$ sheets with four $M$ branch points (d) two $M(z)$ sheets (e) four $M(E)$ sheets and (f) two $E(M)$ sheets. These Riemann sheets are for the model $H_{\text{nh}}$, and the GBZ is calculated from an OBC chain with $N=60$.}
\label{fig:Riemann}
\end{figure}

The band structure equations in Eq.~\eqref{Eq:charpoly} and Eq.~\eqref{Eq:M} can be viewed as Riemann surfaces~\cite{wang2024onedimensional,wang2024nonhermitian,kunst2019nonhermitian,yang2020nonhermitian,hu2023greens,lee2019anatomy,prodan2006analytic,kohn1959analytic,reuter2017unified}. However, the many different projections of these Riemann surfaces, along with the different sets of branch points, can be potentially confusing. We give a glossary of these concepts. The eigenvalue Riemann surface is described by the set of $(E,z) \in \mathbb{C}^2$ given by $f(E,z)=0$. Here, $f(E,z)$ is Eq.~\eqref{Eq:charpoly} converted into a polynomial in both $z$ and $E$. For a general model of Eq.~\eqref{Eq:H} this polynomial has degree 4 in $z$ and degree 2 in $E$. The eigenvector Riemann surface is derived by eliminating $E$ from the two forms of Eq.~\eqref{Eq:M}. This results in an equation that is degree 2 in both $M$ and $z$: 
\begin{align}
g(M, z)=&M^2\left(t_{b a,-1}+z t_{b a, 0}+z^2 t_{b a,+1}\right)-\left(t_{a b,-1}+z t_{a b, 0}+z^2 t_{a b,+1}\right)\nonumber \\
&+M\left(\left(t_{b b,-1}-t_{a a,-1}\right)+z\left(t_{b b, 0}-t_{a a, 0}\right)+z^2\left(t_{b b,+1}-t_{a a,+1}\right)\right).
\label{Eq:MvsZ}
\end{align}
The set of $(M,z) \in \mathbb{C}^2$ satisfying $g(M,z)=0$ gives the eigenvector Riemann surface. Both $g(M, z) = 0$ and $f(E, z) = 0$ define valid Riemann surfaces because the partial derivatives $\frac{\partial g}{\partial M}$, $\frac{\partial g}{\partial z}$ and $\frac{\partial f}{\partial E}$, $\frac{\partial f}{\partial z}$ respectively do not vanish simultaneously in general. There are instances where models of the form given in Eq.~\eqref{Eq:H} fail to describe smooth Riemann surfaces~\cite{wang2024nonhermitian,kawabata2020non}. Such cases are excluded when we refer to the `general' scenario of our model. We can use the above equations to find $h(E,M)=0$, which is a polynomial in both $E$ and $M$. For general models, $h(E, M) = 0$ does not define a smooth Riemann surface, as the partial derivatives $\frac{\partial h}{\partial E}$ and $\frac{\partial h}{\partial M}$ both vanish at the bulk eigenvector degeneracy points. Nevertheless, the visualization of the $M(E)$ and $E(M)$ relations may be useful (SM Sec.~\ref{supp:visualM}). For general cases of Eq.~\eqref{Eq:H}, we have four $z(E)$ sheets~\cite{lee2019anatomy}, two $E(z)$ sheets~\cite{yang2020nonhermitian,hu2023greens}, two $z(M)$ sheets, two $M(z)$ sheets~\cite{zhong2024polezero}, four $M(E)$ sheets and two $E(M)$ sheets, which are illustrated for the model $H_{\text{nh}}$ in Fig.~\ref{fig:Riemann}(a)-(f) respectively. The GBZ curve, $\mathcal{C_{\text{GBZ}}}$, is a 1D curve in $(z,E)$ space defined by
\begin{equation}
    \mathcal{C_{\text{GBZ}}}=\{ (z,E) \text{ where } |z_2|=|z_3| \text{ for Eq.~\eqref{Eq:charpoly}}\}.
\end{equation}
This is plotted in black in Fig.~\ref{fig:Riemann}(a)-(f). The method for solving for the branch points is described in Ref.~\cite{wang2024onedimensional,wang2024nonhermitian}. The $z$-plane, $E$-plane, and $M$-plane branch points are different sets and are plotted in green, blue, and red points respectively in Fig.~\ref{fig:Riemann}(a)-(f). For general cases of our model in Eq.~\eqref{Eq:H}, we have:
\begin{itemize}
\item Four $z$-plane branch points: At each $z$ point, multiple $E$ values coalesce. These are found by calculating the roots of the discriminant of $f(E, z)=0$ with respect to $E$. They are related to OBC gap closing points~\cite{fu2023anatomy} and affect the eigenvalue topology or braid of the GBZ~\cite{fu2024braiding,li2022topological}.
\item Eight $E$-plane branch points:  At each $E$ point, multiple $z$ values coalesce. These are found by calculating the roots of the discriminant of $f(E, z)=0$ with respect to $z$. They are related to `saddle-point energies'~\cite{longhi2019probing,zhou2024abnormal,hu2024geometric,fang2023point} which affects the dynamics of a model~\cite{longhi2019probing,zhou2024abnormal,xue2025non,yang2025real}. Some of the $E$-plane branch points correspond to the end points of the OBC spectra~\cite{wang2024onedimensional,hu2024geometric}. The generalized Bloch ansatz (and therefore Eqs.~\eqref{Eq:C1} to~\eqref{Eq:C4}) does not apply at these $E$-plane branch points, as the mode ansatz takes a modified form~\cite{wang2024nonhermitian}.
\item Four $M$-plane branch points: At each $M$ point, multiple $z$ values coalesce. These are found by calculating the roots of the discriminant of $g(M, z)=0$ with respect to $z$. We use these in our general edge-state invariant in Eq.~\eqref{Eq:generalinvariant}.
\end{itemize}
There are cases where there are less than four $z(E)$ sheets for Eq.~\eqref{Eq:H} such as the SSH model. However, adding a small perturbation of a model with random couplings in Eq.~\eqref{Eq:H} to this model will lead to four $z(E)$ sheets. 

In the main text and in the SM, we used subscripts $i$ with $i=\{1,2,3,4\}$ to denote $z_i(E)$ sheets (and corresponding $M_i(E)=M(z_i,E)$ sheets) and superscripts $(i)$ with $i=\{1,2\}$ to denote $E^{(i)}(z)$ sheets unless otherwise stated. One could label the sheets in any order, but here for $z_i(E)$ sheets (and the $M_i(E)$ sheets corresponding to $z_i(E)$) we use the increasing $|z|$ order convention for labeling. In this context, subscript $i$ can be thought of as one of the terms in the generalized Bloch ansatz, where the Bloch phase factor and wave-vector information is in $z_i$ and the Bloch eigenvector information is in $M_i$. On the other hand, superscript $(i)$ can be thought of as relating to one of the two energy bands.

\subsection{Analytical continuation of wave-vector $k$}
\label{supp:riemannk2k3}
The GBZ~\cite{yao2018edge}, generalized Bloch theorem~\cite{alase2017generalization} and the wave-vectors of edge states all require analytical continuation of the wave-vector $k$ into the complex plane. Here we visualize these concepts. The Riemann surface band structure essentially states that $z=e^{ik}$ for both real and complex $k$ are valid. We can plot all the 4D information $(\text{Re}(E), \text{Im}(E), \text{Re}(k), \text{Im}(k))$ of the band structure for $H_{\text{ssh}}$ and $H_{\text{nh}}$, using color to represent $\text{Re}(k)$ In Fig.~\ref{fig:riemannimk}(a,b). This colored surface and all figures in Fig.~\ref{fig:Riemann} are a visual representation of the generalized Bloch theorem~\cite{alase2017generalization} because the Riemann surface band diagram is the dispersion relation for arbitrary boundary conditions (where the bulk has translational symmetry). Real-space eigenvectors occupy different $E,z$ positions on the colored band diagram surface depending on boundary conditions. For $N=60$ unit cells, we overlay OBC eigenvalues and their $-\log(|z_i|)=\text{Im}(k_i)$ values (found from Eq.~\eqref{Eq:charpoly} and $E$ of the OBC mode) on the band diagram for $H_{\text{ssh}}$ and $H_{\text{nh}}$ in black. For the $H_{\text{ssh}}$ model in Fig.~\ref{fig:riemannimk}(a) where $t_1=0.5,t_2=1$, we have $p=2$ and there are two $z$ solutions per $E$. Bulk OBC states occur where $|z_1|=|z_2|=1$ (plane waves), while the edge state at $E=0$ has unequal $|z_1|\neq|z_2|$ values. For $H_{\text{nh}}$, $p=4$ and there are four $z$ solutions per $E$. We highlight only $z_2$ and $z_3$ in black for Fig.~\ref{fig:riemannimk}(b) for clarity. Here, the GBZ has $|z_2|=|z_3|\neq1$ which are localized modes (this is the `non-Hermitian skin effect'~\cite{gohsrich2024non,lee2016anomalous,yao2018edge,bergholtz2021exceptional,okuma2020topological,zhang2020correspondence}). The GBZ modes may also have varying $|z_2|=|z_3|$ values (and hence localization lengths) for different $E$. The edge state has unequal $|z_2|\neq|z_3|$ values unlike the bulk states. In Fig.~\ref{fig:riemannimk}, we can also see how the GBZ curve sits on certain branch cuts on the Riemann surface~\cite{lee2019anatomy,wang2024nonhermitian,wang2024onedimensional,wu2022connections}.
\begin{figure}[H]
    \centering
    \includegraphics[width=0.65\linewidth]{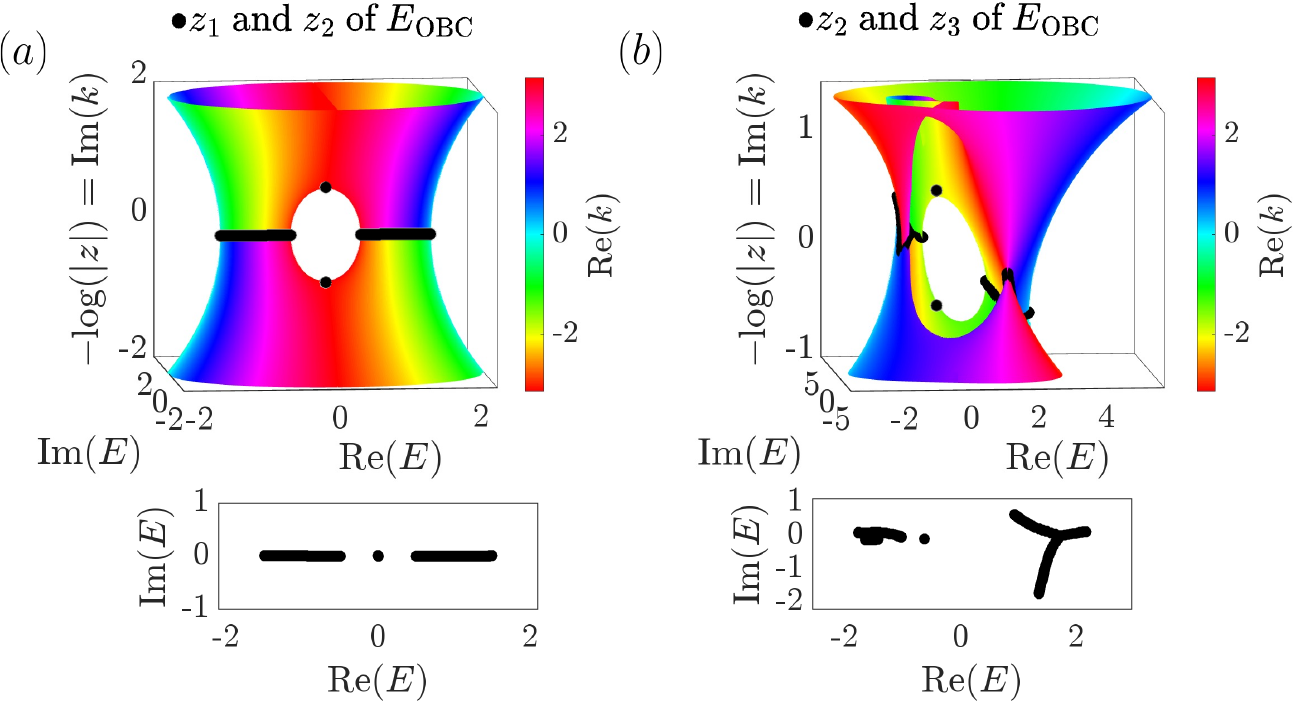}
    \caption{On top, Riemann surface band diagram colored by $\mathrm{Re}(k)$. On bottom, $E_{\text{obc}}$ is plotted on the complex $E$-plane. (a) is the Hermitian SSH for $t_1=0.5, t_1=1$. $z_1$ and $z_2$ for $E_{\text{obc}}$ is plotted in black on top of the band diagram. (b) is the model $H_{\text{nh}}$. $z_2$ and $z_3$ for $E_{\text{obc}}$ is plotted in black on top of the band diagram. }
    \label{fig:riemannimk}
\end{figure}

\section{Bulk eigenvector degeneracy $M_{\text{deg}}$ and $E_{\text{deg}}$ derivation}
\label{supp:MdegEdeg}
In this section, we derive the equation for $M_{\text{deg}}$ in Eq.~\eqref{Eq:Mdeg}. We can write Eq.~\eqref{Eq:M} as 
\begin{equation}
M=\frac{H_{ab}(z)}{E(z)-H_{aa}(z)}=\frac{E(z)-H_{bb}(z)}{H_{ba}(z)}
\end{equation}
where $H$ from Eq.~\eqref{Eq:H} is written as
\begin{equation}
    H= \begin{bmatrix}
        H_{aa} & H_{ab} \\
        H_{ba} & H_{bb}
    \end{bmatrix}.
\end{equation}
This leads to two equations:
\begin{align}
    z[(E - H_{aa})M-H_{ab}]&=0\\
    z[(H_{ba}M-(E-H_{bb})]&=0,
\end{align}
which give:
\begin{align}
\label{Eq:MzE1}
-M t_{a a,-1}-t_{a b,-1}-M t_{a a, 0} z-t_{a b, 0} z-M t_{a a, 1} z^2-t_{a b, 1} z^2+M z E&=0\:, \\
M t_{b a,-1}+t_{b b,-1}+M t_{b a, 0} z+t_{b b, 0} z+M t_{b a, 1} z^2+t_{b b, 1} z^2-z E&=0\:.
\label{Eq:MzE2}
\end{align}
Up to this point, the equations are completely general and can be used to determine $M$ for any pair of $(z,E)$ on the band structure. Now we apply the edge state criterion that at a certain $E$, there are two different $z$ solutions that shares the same $M$ (see discussions around Eqs.~\eqref{Eq:generalinvariant} of the main text as well as the End Matter). This can also be interpreted as having two different $z$ solutions at fixed $E$ and $M$. Since Eq.~\eqref{Eq:MzE1} and \eqref{Eq:MzE2} each has a set of two $z$ solutions and the solution sets are required to be the same, the left hand side of Eqs.~\eqref{Eq:MzE1} and~\eqref{Eq:MzE2} must be proportional to each other. If we then collect the coefficients of $z^2$ and $z^0$ terms, the ratio of the $z^2$ coefficient over the $z^0$ coefficient for both equations must be the same, leading to Eq.~\eqref{Eq:Mdeg}:
\begin{align}
M_{\text{deg}}^2 \left(t_{aa,1} t_{ba,-1} - t_{aa,-1} t_{ba,1}\right) + (t_{ab,1} t_{bb,-1} - t_{ab,-1} t_{bb,1}) & \nonumber\\
+ 
M_{\text{deg}} \left(t_{ab,1} t_{ba,-1} - t_{ab,-1} t_{ba,1} + t_{aa,1} t_{bb,-1} - t_{aa,-1} t_{bb,1}\right) &= 0.
\nonumber
\end{align}
We note that the definition of $M$ and the resulting $M_\text{deg}$ are dependent on the basis for the Hamiltonian; a similarity transformation on $H(z)$ changes the values of $M_\text{deg}$ but not the edge state energies $E_\text{deg}$. As such, we also present an analytical equation for $E_{\text{deg}}$, which can be derived by comparing the coefficients from $z^2$ and $z^1$ terms and substituting $M = M_\text{deg}$. The full expression for $E_{\text{deg}}$ reads:
\begin{align}
E_{\text{deg}} = & \left|\begin{array}{ll}
\langle h_{+} , h_{+} \rangle & \langle h_{+} , h_{-} \rangle \\
\langle h_{-} , h_{+} \rangle & \langle h_{-} , h_{-} \rangle
\end{array}\right|^{-1}\left(-\left|\begin{array}{ccc}
d_{o+} & d_{o0} & d_{o-} \\
\langle h_{+} , h_{+} \rangle & \langle h_{+} , h_{0} \rangle & \langle h_{+} , h_{-} \rangle \\
\langle h_{-} , h_{+} \rangle & \langle h_{-} , h_{0} \rangle & \langle h_{-} , h_{-} \rangle
\end{array}\right|\right. \nonumber \\
& \left.\pm \sqrt{\left|\begin{array}{lll}
\langle h_{+} , h_{+} \rangle & \langle h_{+} , h_{0} \rangle & \langle h_{+} , h_{-} \rangle  \\
\langle h_{0} , h_{+} \rangle & \langle h_{0} , h_{0} \rangle & \langle h_{0} , h_{-} \rangle \\
\langle h_{-} , h_{+} \rangle & \langle h_{-} , h_{0} \rangle & \langle h_{-} , h_{-} \rangle
\end{array}\right|\left(\left|\begin{array}{ll}
\langle h_{+} , h_{+} \rangle & \langle h_{+} , h_{-} \rangle \\
\langle h_{-} , h_{+} \rangle & \langle h_{-} , h_{-} \rangle
\end{array}\right|-\left<d_{o+} h_{-}-d_{o-} h_{+}, d_{o+} h_{-}-d_{o-} h_{+}\right>\right)}\right).
\label{Eq:Edeg}
\end{align}
where $d_{o+} = \operatorname{tr} h_+/2$, $d_{o0} = \operatorname{tr} h_0/2$, $d_{o-} = \operatorname{tr} h_-/2$, and $\left<A,B\right> \equiv \operatorname{tr}[(A - \operatorname{tr} A/2)(B - \operatorname{tr} B/2)]/2$. As the dependence of $E_\text{deg}$ on $H$ is expressed entirely through traces instead of matrix elements, the invariance of $E_\text{deg}$ with respect to similarity transformations have been established. For the nearest-neighbor model considered here, there are only two $E_{\text{deg}}$ values for the entire band structure. For each $E_{\text{deg}}$, four $z$ solutions can be solved from Eq.~\eqref{Eq:charpoly}, corresponding to four $M$ values. Two of these $M$ values becomes  degenerate and equals to $M_{\text{deg}}$. Similar to $M_\text{deg}$, the $E_{\text{deg}}$ values can be calculated regardless of whether $E_{\text{deg}}$ represents an actual edge state. For an $E_{\text{deg}}$ that corresponds to an edge state, the distance of $E_{\text{deg}}$ to the OBC spectrum provides a measure of the stability of the edge state. How far $M_{\text{deg}}$ is to $M_{\text{BZ}}$ or $M_{\text{GBZ}}$ in the Hermitian or non-Hermitian case respectively gives an indication of how far one is to the formation or removal of an edge state (see SM Sec.~\ref{supp:robustness}).

\section{Mode profiles}
\label{supp:modeprofiles}

\begin{figure}[H]
    \centering
    \includegraphics[width=\linewidth]{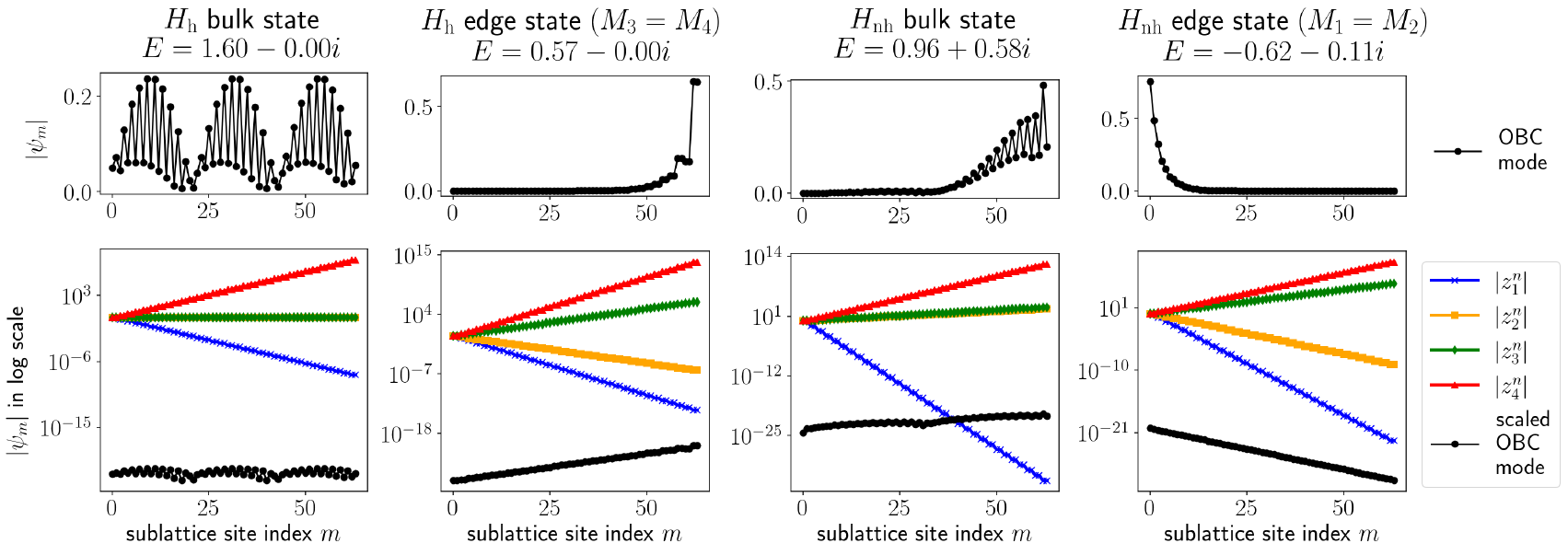}
    \caption{OBC mode profiles of a bulk and edge state for $H_{\text{h}}$ and $H_{\text{nh}}$ where $N=32$ (so there are $2N$ total sublattice sites). On top panel, we plot the numerically diagonalized OBC mode profile from the $2N\times 2N$ matrix using $|\psi_m|$ where $m$ is the sublattice site index. On the bottom row, we plot $\log(|p\psi_m|)$ where $p$ is a constant that shifts the plot along the vertical axis (for visibility) but does not change the gradient (here $p=10^{-20}$). We also solve for $z_1$ to $z_4$ from the $E$ of the numerically diagonalized OBC mode and plot $|z_i^n|$ (where $n=\lfloor \frac{m}{2} \rfloor$ is the unit cell index) to compare the localization direction and lengths (given by the slopes in the bottom plots) of $\psi_m$ to $z_i^n$.}
    \label{fig:modes}
\end{figure}
We give a visualization of mode profiles for bulk and edge OBC states of Hermitian and non-Hermitian models of Eq.~\eqref{Eq:H}. We also show that the localization direction and length information can be obtained from the mode profiles~\cite{lee2019anatomy,wang2024nonhermitian}. If we wanted to understand the localization length and direction of a chain where site $n$ has amplitude $\psi =z^n$, we could plot $\log(|\psi|)$ vs the site index $n$ since $\log(|\psi|)= \log(|z^n|) = n \log(|z|)$ and the gradient of such a plot is $\log(|z|)$.  $\log(|z|)$ contains the localization information of the mode $\psi$ because $\text{Im}(k)=-\log(|z|)$ is the exponential decay or amplification of the mode since $\psi=e^{i \text{Re}(k) n} e^{-\text{Im}(k) n}$. Otherwise, a common measure in literature is localization length~\cite{asboth2015ashort} given by $\xi =-1 / \log \left(\left|z\right|\right)$ in this case. In the $\log(|\psi|)$ vs $n$ plot, if  $|z|=1$ (a plane wave), then $\log(1)=0$ and the gradient is a flat line. If $|z|<1$ (a decaying wave), then $\log(|z|)<0$ and the gradient is negative. If $|z|>1$ (an amplifying wave), then $\log(|z|)>0$ and the gradient is positive.

In general, OBC modes of Eq.~\eqref{Eq:H} have mode amplitude $\left(\psi_{na}, \psi_{nb}\right)^T$ where $n$ is the unit cell index and:
\begin{equation}
    \log(|\psi_{na}|) = \log(|C_1 M_1 z_1^n + C_2 M_2 z_2^n  + C_3 M_3 z_3^n + C_4 M_4 z_4^n|)
\end{equation}
\begin{equation}
    \log(|\psi_{nb}|) = \log(|C_1 z_1^n + C_2 z_2^n  + C_3 z_3^n + C_4 z_4^n|).
\end{equation}
We cannot generally separate out the sums of terms in the logarithm. However, as stated in the main text, for $n=kN$ where $0<k<1$ which are essentially sites in the middle of the chain away from the ends, we can derive from the leading terms in $N\rightarrow\infty$ that for edge states where $M_1=M_2$,
\begin{equation}
    \binom{\psi_{n a}}{\psi_{n b}} \approx C_2 z_2^n\binom{M_2}{1}.
\end{equation}
Then in the large $N$ limit, $\log |\psi_{n\mu}| \propto n \log |z_2|$ for $\mu=a$ or $b$. Similarly, for edge states where $M_3=M_4$ in $N \rightarrow \infty$,
\begin{equation}
    \binom{\psi_{n a}}{\psi_{n b}} \approx C_3 z_3^n\binom{M_3}{1}.
\end{equation}
and in the large $N$ limit, $\log |\psi_{n\mu}| \propto n \log |z_3|$ for $\mu=a$ or $b$. For bulk states, $|z_2|=|z_3|$ in $N \rightarrow \infty$,
\begin{equation}
    \binom{\psi_{n a}}{\psi_{n b}} \approx \sum_{i=2}^3 C_i z_i^n\binom{M_i}{1}
\end{equation}
and in the large $N$ limit, the envelope of the amplitude of $\log |\psi_{n\mu}|$ is $ \propto n \log |z_2| = n \log |z_3|$ for $\mu=a$ or $b$. As the bulk states are a sum of two $z_i^n$ terms, the phases of the $z_i$ terms can cause interference and oscillations, but their decay length is still given by the aforementioned condition.

We visualize this information for an $N=32$ edge and bulk mode profile for the Hermitian model $H_{\text{h}}$ and the non-Hermitian model $H_{\text{nh}}$ in Fig.~\ref{fig:modes}. Let $\psi_m$ be the $m$-th component of the $2N\times1$ real-space OBC eigenvector $\psi$ of eigenvalue $E$ found from numerically diagonalizing the $2N\times 2N$ real-space matrix, where we index by sublattice site $m$ rather than unit cell $n$. In the top panels of Fig.~\ref{fig:modes}, we plot the magnitude of the diagonalized OBC mode profile.  In the bottom panels of Fig.~\ref{fig:modes}, we again plot the magnitude of the diagonalized OBC mode profile but in log scale as $\log(|p\psi_m|)$ where $p$ is a constant that shifts the plot along the vertical axis (for visibility) but does not change the gradient. Here $p=10^{-20}$. We also solve for the four $z(E)$ values from the OBC mode eigenvalue $E$ using Eq.~\eqref{Eq:charpoly} and plot $|z_i^n|$ in log scale where $n=\lfloor \frac{m}{2} \rfloor$ is the unit cell index. The gradient of these $|z_i^n|$ plots is $\log(|z_i|)$. We use this to compare the localization direction and lengths (given by the gradients) of $\psi_m$ to the $z_i^n$ components. We see that for the bulk states, $\log(|\psi_m|)$ has a slope approximately matching $\log(|z_2|^n)$ and $\log(|z_3|^n)$ which verifies $\binom{\psi_{n a}}{\psi_{n b}} \approx \sum_{i=2}^3 C_i z_i^n\binom{M_i}{1}$. For the $M_1=M_2$ edge state, $\log(|\psi_m|)$ has a slope approximately matching $\log(|z_2|^n)$ which verifies $\binom{\psi_{n a}}{\psi_{n b}} \approx C_2 z_2^n\binom{M_2}{1}$.  For the $M_3=M_4$ edge state, $\log(|\psi_m|)$ has a slope approximately matching $\log(|z_3|^n)$ which verifies $\binom{\psi_{n a}}{\psi_{n b}} \approx C_3 z_3^n\binom{M_3}{1}$.

\section{Proof of the general edge-state invariant}
\label{supp:generalinvariant}

The proof consists of several components. In SM Sec. \ref{supp:generalinvariant_RS} we show that, under general conditions, the band structure defines a Riemann surface that is topologically equivalent to a torus. This torus can be represented by the Weierstrass elliptic function $\wp(\tau; g_2,g_3)$, where $\tau$ is the ``torus parameter'' and the elliptic invariants $g_2$ and $g_3$ can be obtained from the coupling coefficients. Then in SM Sec. \ref{supp:generalinvariant_WN} we formulate the edge state condition as a winding integral on the $\tau$ plane. With a change of variables from $\tau$ to $M$, the integration can be interpreted as a winding number on the $M$ plane. In SM Sec. \ref{supp:generalinvariant_DS} we show the additional property that the two edge states cannot share the same side of the chain, which is not necessary for the proof but can be applied to specific cases.

\subsection{Riemann surface for the band structure}
\label{supp:generalinvariant_RS}
We begin with the band structure $\operatorname{det}(H-E\mathbb{I})=0$, which can be expanded as
\begin{equation}
E^2 - E \operatorname{tr}H + \operatorname{det}H = 0
\end{equation}
The discriminant of the band structure with respect to $E$, denoted as $\Delta$, can be found as
\begin{equation}
\Delta = (\operatorname{tr}H)^2 - 4 \operatorname{det}H = - \operatorname{det}(2H-\operatorname{tr}H)
\label{Eq:discz}
\end{equation}
and the band structure can be rearranged as
\begin{equation}
z^2\left(E - \frac{\operatorname{tr}H}{2}\right)^2 = \frac{z^2\Delta}{4}
\label{Eq:elliptic_band}
\end{equation}
Under the assumptions that $H$ only contains nearest-unit-cell couplings, each entry of $H$ is a Laurent polynomial in $z$ with powers of $z$ from $-1$ to $1$, and $z^2\Delta$ is a polynomial in $z$ with a degree at most four. The fundamental theorem of algebra implies that $z^2\Delta=0$ then has four solutions (if the degree of $z^2\Delta$ is less than four, we include $z = \infty$ in the solution list).

The band structure, after multiplication by $z^2$, is a polynomial in $z$ and $E$, and therefore is an algebraic curve. An algebraic curve is also a compact Riemann surface if the curve is smooth. For the band structure, the smoothness condition is that there are no duplicate roots in $z^2\Delta=0$ (including $z = \infty$). This is readily satisfied by models with general coefficients. We will assume that $z^2\Delta=0$ does not contain duplicate roots from here on. In such cases, Eq.~\eqref{Eq:elliptic_band} represents an elliptic curve of genus $1$ and its Riemann surface is topologically equivalent to a torus~\cite{beals2020explorations}.

We now write $\Delta = \delta_{-2} z^{-2} + \delta_{-1} z^{-1} + \delta_0 + \delta_1 z + \delta_2 z^2$, where the $\delta$'s are complex numbers determined from $H$. We also denote one of the finite roots of $z^2\Delta = 0$ as $z_\Delta$ and define the following quantities:
\begin{equation}
g_2 = 16\delta_{-2}\delta_2 - 4\delta_{-1}\delta_1 + \frac{4}{3}\delta_0^2\:,
\end{equation}
\begin{equation}
g_3 = -4\delta_{-2}\delta_1^2 -4\delta_{-1}^2\delta_2 + \frac{32}{3}\delta_{-2}\delta_0\delta_2 + \frac{4}{3}\delta_{-1}\delta_0\delta_1 -\frac{8}{27}\delta_0^3\:,
\end{equation}
\begin{equation}
p_1 = \delta_{-1} + 2\delta_0 z_\Delta + 3\delta_1 z_\Delta^2 + 4\delta_2 z_\Delta^3\:,
\end{equation}
\begin{equation}
p_2 = \frac{1}{3}(\delta_0 + 3\delta_1 z_\Delta + 6 \delta_2 z_\Delta^2)\:.
\end{equation}
Then the following parametrization expresses $z$ and $E$ in terms of the Weierstrass elliptic function $\wp(\tau; g_2,g_3)$:
\begin{equation}
z = z_\Delta + \frac{p_1}{\wp(\tau; g_2,g_3) - p_2}
\end{equation}
\begin{equation}
E = \frac{\operatorname{tr}H}{2} + \frac{\wp'(\tau; g_2,g_3)}{4z}\frac{p_1}{[\wp(\tau; g_2,g_3) - p_2]^2}
\end{equation}
where $\wp(\tau; g_2,g_3)$ satisfies the differential equation $\wp'^2 = 4\wp^3 - g_2 \wp - g_3$ and the prime indicates derivative with respect to $\tau$.

The Weierstrass elliptic function is doubly periodic in the complex $\tau$ plane, i.e. $\wp(\tau + \omega_1; g_2,g_3) = \wp(\tau + \omega_2; g_2,g_3) = \wp(\tau; g_2,g_3)$, where the periods $\omega_1$ and $\omega_2$ are linearly-independent complex numbers as functions of $g_2$ and $g_3$. As such, $\tau$, $\tau+\omega_1$, and $\tau+\omega_2$ all represent the same $(z,E)$ point on the Riemann surface. It is sufficient to consider the range of $\tau$ within the fundamental parallelogram bounded by $0$, $\omega_1$, $\omega_1+\omega_2$, and $\omega_2$ on the $\tau$ plane (see Fig.~\ref{fig:weierstrass} for an example), where each $(z,E)$ has a one-to-one correspondence with $\tau$. In addition, opposite edges of this parallelogram should be ``glued together'' (i.e., identified) due to periodicity. This shows that the Riemann surface, parametrized by $\tau$, is topologically equivalent to a torus.

\begin{figure}[H]
\centering
\includegraphics[width=\textwidth]{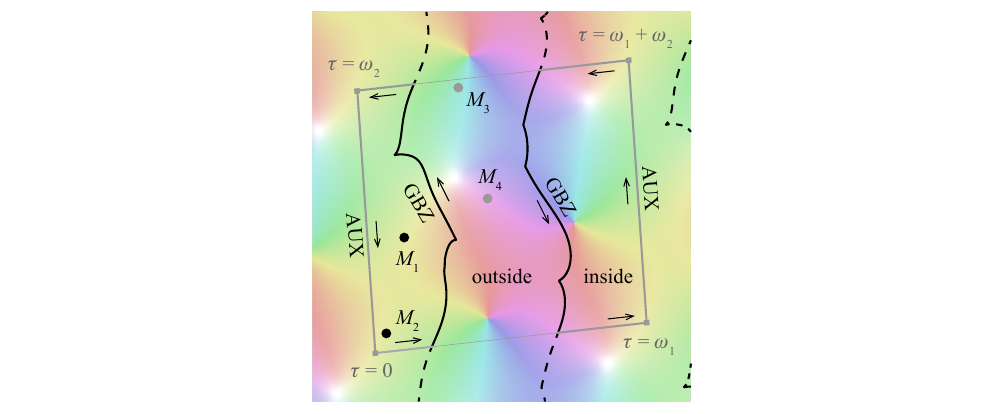}
\caption{The complex $M$ as a function of the ``torus parameter'' $\tau$. Here the Hamiltonian is given by $H_\text{nh}$ (see End Matter and Fig.~\ref{fig:numericalstepsNH}). The $\tau$ plane is colored by $M$ values where the color gives $\arg(M)$ and the brightness gives $|M|$. The fundamental parallelogram (gray) is bounded by $\tau=0$, $\tau=\omega_1$, $\tau=\omega_1+\omega_2$, and $\tau=\omega_2$. The set of $M$ values for the only edge state in the model, where $M_1 = M_2$, have been marked on the plane as dots. The GBZ (black curves) extends infinitely on the $\tau$ plane, but only the sections inside the fundamental parallelogram (solid) are being considered here. The GBZ divides the fundamental parallelogram into an outside and an inside. The integration contour consists of the GBZ itself as well as auxiliary contours along the parallelogram (thick gray lines). The contour direction (arrows) is chosen such that it circles around the GBZ inside in the counterclockwise direction.
}
\label{fig:weierstrass}
\end{figure}

\subsection{Number of edge states as a contour integration}
\label{supp:generalinvariant_WN}
In the End Matter, we have proved that the existence of edge states requires $M_1 = M_2$ or $M_3 = M_4$. Before formulating this condition as an integral on the GBZ, we require some properties of the mapping from the Riemann surface of the band structure to $M$, which we discuss in the following.

We first show that there are exactly two zeros of $M(z,E) - M_\text{deg}$ on the torus, counted with multiplicity. This can be seen from Eq.~\eqref{Eq:MvsZ}, which is a quadratic equation of $z$. In fact, this conclusion is independent of the edge state criterion, and $M_\text{deg}$ can be replaced with any complex number. An arbitrary value of $M$ has two preimages on the torus if $M$ is not a branch point. If $M$ is a branch point, the preimage is a single point on the torus but the zero of $M - M_\text{branch}$ has multiplicity two.

Now we introduce the notion of the inside and outside of the GBZ on the band structure Riemann surface. We define a point $(z,E)$ to be inside the GBZ if $z$ is among the two roots with the smallest amplitudes ($z=z_1$ or $z=z_2$, where $z_1$ to $z_4$ are the four solutions sorted in non-decreasing amplitude of the band structure equation at $E$). By the continuity of polynomial roots, all points inside the GBZ form (possibly disconnected) regions and are separated from the outside of the GBZ by the GBZ $|z_2| = |z_3|$. The edge state criterion can then be stated as whether the preimages of a single $M_\text{deg}$ are both inside or both outside the GBZ on the Riemann surface. If both preimages are inside the GBZ, then $M_\text{deg} = M_1 = M_2$ and corresponds to an edge state. Similarly, if both preimages are outside the GBZ, then $M_\text{deg} = M_3 = M_4$ and also corresponds to an edge state. If one preimages is inside the GBZ and the other is outside the GBZ, then $M_\text{deg}$ does not correspond to an edge state. The number of preimages inside the GBZ for $M_\text{deg}$ can also be viewed as the number of times the region containing $M_\text{deg}$ is covered by the mapping from the inside of the GBZ to the $M$-Riemann sphere.

We can now assign a contour direction to each piece of the GBZ on the torus (Fig.~\ref{fig:weierstrass}), where the inside of the GBZ on the $\tau$ plane appears on the left when moving in the positive direction of the contour. We also add auxiliary contours on the boundaries of the fundamental parallelogram (denoted as AUX) so that the inside of the GBZ is completely enclosed in the contour. We can then define the winding integral of $M - M_\text{deg,1}$ along the boundary of the inside of the GBZ:
\begin{equation}
w_\text{deg,1} = \frac{1}{2\pi i}\oint_\text{GBZ + AUX} d\tau \frac{d}{d\tau} \ln (M(\tau) - M_\text{deg,1})
\end{equation}
The subscript $1$ indicates that only one of the $M_\text{deg}$ is currently being considered.
By the argument principle, this integral will count the number of zeros minus the number of poles of $M - M_\text{deg,1}$ inside the GBZ. We note that the pole of $M - M_\text{deg,1}$ is simply the pole of $M$. To remove this quantity, we consider the winding integral of $M - M_\text{branch}$, where $M_\text{branch}$ is any one of the four branch points on the $M$ sphere:
\begin{equation}
w_\text{branch} = \frac{1}{2\pi i}\oint_\text{GBZ + AUX} d\tau \frac{d}{d\tau} \ln (M(\tau) - M_\text{branch})
\end{equation}
The difference of these two winding numbers give the difference of the number of zeros for $M - M_\text{deg,1}$ and $M - M_\text{branch}$. The numbers of zeros for $M - M_\text{deg}$ is even if $M_\text{deg,1}$ leads to an edge state, and odd if not an edge state. On the other hand, the number of zeros of $M - M_\text{branch}$ is always an even number due to the degeneracy. Therefore, we can define
\begin{equation}
w_1 = w_\text{deg,1} - w_\text{branch} = \frac{1}{2\pi i}\oint_\text{GBZ + AUX} d\tau \frac{d}{d\tau} \ln \frac{M(\tau) - M_\text{deg,1}}{M(\tau) - M_\text{branch}} \bmod 2
\end{equation}
The $\bmod$ $2$ operation effectively removes the contributions from zeros of $M - M_\text{branch}$. As such, $w_1 = 0$ indicates an edge state from $M_\text{deg,1}$, while $w_1 = 1$ indicates no edge states from $M_\text{deg,1}$. Another difference $w_2 = w_\text{deg,2} - w_\text{branch}$ can be defined for $M_\text{deg,2}$.

Finally, we note that the integrations on the auxiliary contours will cancel each other due to the periodicity of $\tau$, leading to
\begin{equation}
w_1 = \frac{1}{2\pi i}\oint_\text{GBZ} d\tau \frac{d}{d\tau} \ln \frac{M(\tau) - M_\text{deg,1}}{M(\tau) - M_\text{branch}} \bmod 2
\end{equation}
A change of variables from $\tau$ to $M$ then converts the integral to the $M$ plane:
\begin{equation}
w_1 = \frac{1}{2\pi i}\oint_{M(\mathcal{C}_\text{GBZ})} dM \frac{d}{dM} \ln \frac{M - M_\text{deg,1}}{M - M_\text{branch}} \bmod 2
\end{equation}
We also note that reversing the orientation of any loop will change the winding numbers by an even number, so the orientation of GBZ is not required as long as the integration paths forms loops along the GBZ:
\begin{equation}
w_1 = \frac{1}{2\pi i}\int_{M(\mathcal{C}_\text{GBZ})} dM \frac{d}{dM} \ln \frac{M - M_\text{deg,1}}{M - M_\text{branch}} \bmod 2
\end{equation}
An even $w_1$ indicates an edge state. Adding the parity flip by defining $W_1 = w_1 + 1 \bmod 2$ reproduces Eq.~\eqref{Eq:generalinvariant} and completes the proof.

\subsection{Distributions of the edge states}
\label{supp:generalinvariant_DS}
In this subsection we prove that the two edge states, if both are present, cannot be all on the left edge or all on the right edge. Here ``on the left edge'' is defined by $M_1 = M_2$ and ``on the right edge'' is defined by $M_3 = M_4$. This is consistent with the fact that the $M_1 = M_2$ condition originates from the left boundary regardless of the actual direction of localization of the state. Although this knowledge is not necessary for the proof of the winding numbers, it can be used to further simplify the integral in some special cases such as in Hermitian models. The idea is that the states participating in edge state formations cannot all have small $|z|$ amplitudes and therefore must belong to different edges.

Denote $z_\text{A1}$ and $z_\text{A2}$ as solutions to Eqs.~\eqref{Eq:MzE1} and \eqref{Eq:MzE2} when $M =  M_\text{deg,1}$, and $z_\text{B1}$ and $z_\text{B2}$ when $M =  M_\text{deg,2}$. Applying Vieta's formula to Eq.~\eqref{Eq:MzE1} leads to
\begin{equation}
|z_\text{A1} z_\text{A2}| = \left|\frac{M_\text{deg,1} t_{a a,-1} + t_{a b,-1}}{M_\text{deg,1} t_{a a, 1} + t_{a b, 1}}\right|
\end{equation}
\begin{equation}
|z_\text{B1} z_\text{B2}| = \left|\frac{M_\text{deg,2} t_{a a,-1} + t_{a b,-1}}{M_\text{deg,2} t_{a a, 1} + t_{a b, 1}}\right|
\end{equation}
These can be multiplied together and results in
\begin{equation}
|z_\text{A1} z_\text{A2} z_\text{B1} z_\text{B2}| = \left|\frac
{t_{a a,-1}^2 M_\text{deg,1} M_\text{deg,2} + t_{a a,-1} t_{a b,-1} (M_\text{deg,1} + M_\text{deg,2}) + t_{a b,-1}^2}
{t_{a a, 1}^2 M_\text{deg,1} M_\text{deg,2} + t_{a a, 1} t_{a b, 1} (M_\text{deg,1} + M_\text{deg,2}) + t_{a b, 1}^2}\right|
\end{equation}
The values of the symmetric polynomials $M_\text{deg,1} M_\text{deg,2}$ and $M_\text{deg,1} + M_\text{deg,2}$ can be obtained by applying Vieta's formula to Eq.~\eqref{Eq:Mdeg}. After some rearrangements we get
\begin{equation}
|z_\text{A1} z_\text{A2} z_\text{B1} z_\text{B2}| = \left|\frac{t_{a a,-1} t_{b b,-1} - t_{a b,-1} t_{b a,-1}}{t_{a a, 1} t_{b b, 1} - t_{a b, 1} t_{b a, 1}}\right| = \left|\frac{\operatorname{det}h_-}{\operatorname{det}h_+}\right|
\label{Eq:edge_z_balance}
\end{equation}

On the other hand, we can expand the band structure $E^2 - E \operatorname{tr}H + \operatorname{det}H = 0$, which results in
\begin{equation}
z^{-2}\operatorname{det}h_- + \cdots + z^{2}\operatorname{det}h_+ = 0
\end{equation}
where the $z^{-1}$, $z^0$ and $z^1$ terms are not relevant to the current analysis and are therefore omitted. At any $E$, application of the Vieta's formula to the band structure leads to
\begin{equation}
|z_1 z_2 z_3 z_4| = \left|\frac{\operatorname{det}h_-}{\operatorname{det}h_+}\right|
\end{equation}
Now assume that both edges states are on the left edge, i.e. $z_\text{A1}$ and $z_\text{A2}$ are the two solutions with the smallest magnitude among the four $z$ solutions of $E_\text{deg,1}$ and the same holds for $E_\text{deg,2}$. This would imply that
\begin{equation}
|z_\text{A1} z_\text{A2}| < \left|\frac{\operatorname{det}h_-}{\operatorname{det}h_+}\right|^{1/2}
\end{equation}
\begin{equation}
|z_\text{B1} z_\text{B2}| < \left|\frac{\operatorname{det}h_-}{\operatorname{det}h_+}\right|^{1/2}
\end{equation}
These can be multiplied together and results in
\begin{equation}
|z_\text{A1} z_\text{A2} z_\text{B1} z_\text{B2}| < \left|\frac{\operatorname{det}h_-}{\operatorname{det}h_+}\right|
\end{equation}
which is in direct contradiction to Eq.~\eqref{Eq:edge_z_balance}. We therefore conclude that if there are two edge states in the same system, one must be on the left edge and the other one must be on the right edge.

\section{Proof of the Hermitian edge-state invariant}
\label{supp:restrictedinvariant}

Here we consider the special case of Hermitian models. We show that under a quasi-chiral basis as defined below, the eigenmodes from one band will always be concentrated on one of the lattice sites. Together with the orthogonality of modes (inversion symmetry of $M$) between the two bands, this implies that the two $M(\mathcal{C}_\text{BZ})$ loops are simple loops on the $M$-Riemann sphere. We also use the fact that if there exists two edge states, they ca not originate from the same side of the chain (SM Sec. \ref{supp:generalinvariant_DS}). These results can be combined to remove the $M_\text{branch}$ correction from the general winding number and lead to integrations over each band that more closely resemble the invariants of the Berry-Zak phase as applicable to models with symmetries.

For later convenience we rewrite the Hamiltonian as
\begin{equation}
H = \frac{\operatorname{Tr}H}{2}\mathbb{I} + \left(\mathbf{v}_0 + \mathbf{v}_\text{sym} \cos k + \mathbf{v}_\text{asym} \sin k \right) \cdot \bm{\sigma}^*
\end{equation}
where $\mathbf{v}_0 = \operatorname{Tr}(h_0\bm{\sigma}^*)/2$, $\mathbf{v}_\text{sym} = \operatorname{Tr}[(h_++h_-)\bm{\sigma}^*]/2$ and $\mathbf{v}_\text{asym} = i\operatorname{Tr}[(h_+-h_-)\bm{\sigma}^*]/2$ are three real 3D vectors and $\bm{\sigma} = (\sigma_x, \sigma_y, \sigma_z)$ are the vector of Pauli matrices. The conjugation for $\bm{\sigma}$ replaces $\sigma_y$ with $-\sigma_y$ and aligns the Bloch-sphere coordinates with the $M$ Riemann sphere defined here. Applying unitary transformations to $H$ is equivalent to simultaneously applying the same rotation to $\mathbf{v}_0$, $\mathbf{v}_\text{sym}$ and $\mathbf{v}_\text{asym}$. As such, we can choose a basis where both $\mathbf{v}_\text{sym}$ and $\mathbf{v}_\text{asym}$ have no $Z$ components. This is achieved by setting the $Z$ direction perpendicular to the plane formed by $\mathbf{v}_\text{sym}$ and $\mathbf{v}_\text{asym}$. We term this basis a quasi-chiral basis. 

Using Eq.~\eqref{Eq:MRiemann}, the location of eigenvectors on the $M$ Riemann sphere is given by the direction of $\pm (\mathbf{v}_0 + \mathbf{v}_\text{sym} \cos k + \mathbf{v}_\text{asym} \sin k)$. Under the quasi-chiral basis, the $Z$ component of each band is $\mathbf{v}_{0,Z}$ and is a constant. For $\mathbf{v}_{0,Z} \neq 0$, one band is restricted to the southern hemisphere and the other band is restricted to the northern hemisphere. In addition, since the vector $\mathbf{v}_0 + \mathbf{v}_\text{sym} \cos k + \mathbf{v}_\text{asym} \sin k$ represents an ellipse in the 3D space, its Bloch sphere projection will divide the sphere into at most two regions (division does not occur if the projected ellipse degenerates into a backtracking curve). Using the correspondence between the Bloch sphere and the $M$-Riemann sphere, both $M(\mathcal{C}_\text{BZ})$ loops do not cross the $|M| = 1$ circle and divides the $M$-Riemann sphere into at most three regions [Fig.~\ref{fig:combinatoricH}(a)]: the region having the $M(\mathcal{C}_\text{BZ})$ loop with $|M| < 1$ as its boundary (designated as Region I), the region containing $|M| = 1$ (Region II), and the region having the $M(\mathcal{C}_\text{BZ})$ loop with $|M| > 1$ as its boundary (Region III). For $\mathbf{v}_{0,Z} = 0$, the $M(\mathcal{C}_\text{BZ})$ is restricted to the $|M| = 1$ circle, and divides the $M$-Riemann sphere into a single Region II or two regions (I and III), depending on whether the loop circles around the origin. We note that the two $M(\mathcal{C}_\text{BZ})$ are central symmetric with respect to the Riemann sphere center due to the orthogonal relation between the eigenstates at each wavevector.

The Hermiticity of the model imposes the constraints $h_0 = h_0^\dagger$ and $h_+ = h_-^\dagger$ on the coupling coefficients. By comparing the eigenstates of $H$ and $H^\dagger$ at complex $z$ values, $-1/M^*$ and $1/z^*$ are solutions to Eqs.~\eqref{Eq:MzE1} and \eqref{Eq:MzE2} if $M$ and $z$ are solutions. As the BZ is simply the unit circle $|z| = 1$, the number of zeros for $M - M_\text{deg}$ inside the BZ equals the number of zeros for $M + 1/M_\text{deg}^*$ outside the BZ. In addition, the number of zeros inside the BZ remains the same for each $M_\text{deg}$ inside the same region. If $M_\text{deg}$ belongs to Region II, then $-1/M_\text{deg}^*$ also belongs to Region II, and as such has exactly a single zero inside the BZ. Similarly, the number of zeros inside the BZ for regions I and III adds up to two. Since they are across the $M(\mathcal{C}_\text{BZ})$ loop from Region II, they cannot each have a single zero in the BZ. As such, either Region I has no zeros inside the BZ and Region III has two, or vice versa.

\begin{figure}[H]
\centering
\includegraphics[width=\textwidth]{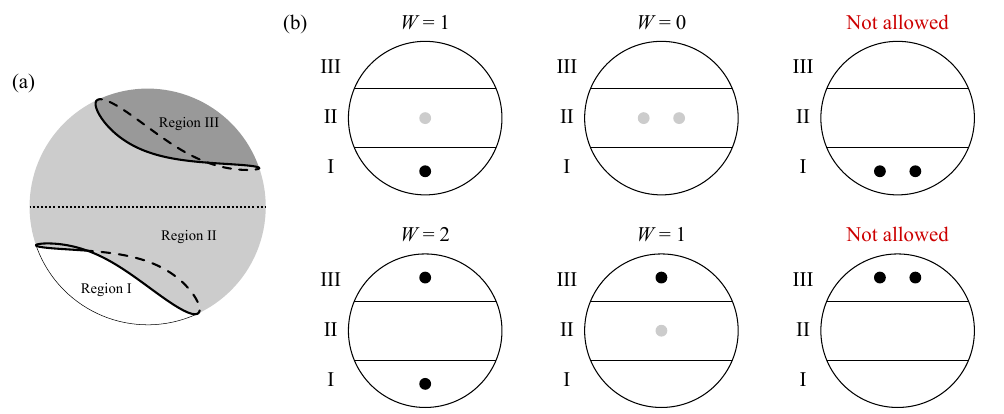}
\caption{The $M$ Riemann sphere in the quasi-chiral basis and the combinatorial argument for the Hermitian invariant.
(a) The $M$ Riemann sphere for the Hamiltonian $H_\text{h}$ (see End Matter for parameters and Fig.~\ref{fig:numericalstepsH}). The viewing direction is the same as Fig.~\ref{fig:proof}(b) and is already in a quasi-chiral basis with $Z$ pointing up and $X$ pointing right. The two $M(\mathcal{C}_\text{BZ})$ loops (black curves) are separated by the equator ($|M| = 1$, dotted) and divides the sphere into Regions I, II and III. Region II indicates a single zero inside the BZ (light gray) and Region III indicates two zeros (dark gray).
(b) All possible $M_\text{deg}$ (dots) locations in the three regions on the $M$ Riemann sphere. Black dots indicate the $M_\text{deg}$ points are edge states, and gray dots indicate the $M_\text{deg}$ points are not states. For the first four configurations, the results of Eq.~\eqref{Eq:restricted} and Eq.~\eqref{Eq:generalinvariant} agree, while the last two configurations are not allowed (SM Sec.~\ref{supp:generalinvariant_DS}).
}
\label{fig:combinatoricH}
\end{figure}

At this point, the proof becomes a combinatorial problem of listing all the assignments of $M_\text{deg,1}$ and $M_\text{deg,2}$ to the three regions [Fig.~\ref{fig:combinatoricH}(b)]. A $M_\text{deg}$ point in regions I or III indicates an edge state due to the even number of zeros inside the BZ, while a $M_\text{deg}$ in Region II is not an edge state. In addition, the configurations where the two points are both in Region I or Region III are not allowed, as this would indicate two edge states on the same edge. This leaves four possibilities: the points may be in Regions I and II, I and III, II and II, or II and III. In each case the Hermitian integral agrees with the general integral, and its validity can thus be established in the quasi-chiral basis.

Finally, we note that Eq.~\eqref{Eq:restricted} is invariant under Möbius transformations of $M$. As the unitary transformation performs rotations on the $M$-Riemann sphere, which is a subset of the Möbius transformations, the integral remains valid and is independent of the basis.

This proof does not generalize to non-Hermitian systems because the $M(\mathcal{C}_\text{GBZ})$ loops may intersect with each other on the $M$ sphere and create more than three regions.

\section{Longer-range models}
\label{supp:secondnearest}
We discuss the generalization of concepts in our main text to longer-range models. First we provide an argument for the topological criteria for the existence edge states in a tight-binding model with arbitrary couplings and on-site potentials for second-nearest neighbor coupling between unit cells. Such a model has a Bloch matrix given by:
\begin{equation}
H(z)=\left(\begin{array}{cc}
\frac{t_{a a,-2}}{z^2}+\frac{t_{a a,-1}}{z}+t_{a a, 0}+t_{a a, 1} z+t_{a a, 2} z^2 & \frac{t_{a b,-2}}{z^2}+\frac{t_{a b,-1}}{z}+t_{a b, 0}+t_{a b, 1} z+t_{a b, 2} z^2 \\
\frac{t_{b a,-2}}{z^2}+\frac{t_{b a,-1}}{z}+t_{b a, 0}+t_{b a, 1} z+t_{b a, 2} z^2 & \frac{t_{b b,-2}}{z^2}+\frac{t_{b b,-1}}{z}+t_{b b, 0}+t_{b b, 1} z+t_{b b, 2} z^2
\end{array}\right).
\label{Eq:NNNmodel}
\end{equation}
The characteristic polynomial of this model gives eight $z$ solutions for each $E$ and two $E$ solutions for each $z$. Under OBC, the real-space equations from the boundary sites give eight equations. Using the generalized Bloch ansatz for this model (which is now a sum of eight terms), we can derive a $8\times8$ coefficient matrix similar to that in the End Matter. The determinant from the coefficient matrix must be zero for nontrivial OBC solutions which gives
\begin{equation}
\left|\begin{array}{cccccccc}
M_1 z_1^{-1} & M_2 z_2^{-1} & M_3 z_3^{-1} & M_4 z_4^{-1} & M_5 z_5^{-1} & M_6 z_6^{-1} & M_7 z_7^{-1} & M_8 z_8^{-1} \\
z_1^{-1} & z_2^{-1} & z_3^{-1} & z_4^{-1} & z_5^{-1} & z_6^{-1} & z_7^{-1} & z_8^{-1} \\
M_1 & M_2 & M_3 & M_4 & M_5 & M_6 & M_7 & M_8 \\
1 & 1 & 1 & 1 & 1 & 1 & 1 & 1 \\
M_1 z_1^N & M_2 z_2^N & M_3 z_3^N & M_4 z_4^N & M_5 z_5^N & M_6 z_6^N & M_7 z_7^N & M_8 z_8^N \\
z_1^N & z_2^N & z_3^N & z_4^N & z_5^N & z_6^N & z_7^N & z_8^N \\
M_1 z_1^{N+1} & M_2 z_2^{N+1} & M_3 z_3^{N+1} & M_4 z_4^{N+1} & M_5 z_5^{N+1} & M_6 z_6^{N+1} & M_7 z_7^{N+1} & M_8 z_8^{N+1} \\
z_1^{N+1} & z_2^{N+1} & z_3^{N+1} & z_4^{N+1} & z_5^{N+1} & z_6^{N+1} & z_7^{N+1} & z_8^{N+1}
\end{array}\right|=0.
\end{equation}
The leading term of this determinant is given by 
\begin{equation}
    \left(\frac{1}{z_1 z_2 z_3 z_4} \times \mathcal{D}_{\text{left}} \times \mathcal{D}_{\text{right}}\right) z_5^N z_6^N z_7^N z_8^N
\end{equation}
where 
\begin{equation}
\mathcal{D}_{\text{left}}=\left|\begin{array}{cccc}
M_1 & M_2 & M_3 & M_4 \\
1 & 1 & 1 & 1 \\
M_1 z_1 & M_2 z_2 & M_3 z_3 & M_4 z_4 \\
z_1 & z_2 & z_3 & z_4
\end{array}\right|, \quad
\mathcal{D}_{\text{right}}=\left|\begin{array}{cccc}
M_5 & M_6 & M_7 & M_8 \\
1 & 1 & 1 & 1 \\
M_5 z_5 & M_6 z_6 & M_7 z_7 & M_8 z_8 \\
z_5 & z_6 & z_7 & z_8
\end{array}\right|.
\end{equation}
Thus, the analytical edge state condition is given by 
\begin{equation}
    \mathcal{D}_{\text{left}}=0\text{ or }\mathcal{D}_{\text{right}}=0.
    \label{Eq:NNNedgecondition}
\end{equation}
The second leading term is proportional to $z_4^N z_6^N z_7^N z_8^N$, and the other feasible solution in the $N \rightarrow \infty$ limit is $|z_4|=|z_5|$, which are the bulk states formed by the GBZ~\cite{yokomizo2019nonbloch}. We note that the edge state criteria in Eq.~\eqref{Eq:NNNedgecondition} is no longer the simple $M_1=M_2$ or $M_3=M_4$ condition from the nearest-neighbor case. Instead, $\mathcal{D}_{\text{left}}$ and $\mathcal{D}_{\text{right}}$ contains explicit $z$-dependence. Furthermore, we do not currently have a simple form for the $E$ at which $\mathcal{D}_{\text{left}}=0$ or $\mathcal{D}_{\text{right}}=0$ occurs. However, one can still make an argument for the topological nature of edge states in asymmetric next-nearest-neighbor models. We note that the quantities

\begin{align}
\mathcal{D}_{ijkl} = 
\left|\begin{array}{cccc}
M_i & M_j & M_k & M_l \\
1 & 1 & 1 & 1 \\
M_i z_i & M_j z_j & M_k z_k & M_l z_l \\
z_i & z_j & z_k & z_l
\end{array}\right| \quad \text{ where } i, j ,k,l\in\{1,2,3,4,5,6,7,8\} \text{ and all distinct}
\label{Eq:NNNdeterminant}
\end{align}
are rational functions of $M_i$ and $z_i$, and are therefore roots to a polynomial equation expressible in terms of the coupling coefficients. Although the condition for potential edge states in second nearest-neighbor models is slightly different from the nearest neighbor case, we continue to use the term `bulk eigenvector degeneracy points' to describe the solutions of Eq.~\eqref{Eq:NNNdeterminant} as it indicates a linear dependency of vectors formed from bulk quantities. Since Eq.~\eqref{Eq:NNNdeterminant} is algebraic, by B\'ezout's theorem, there is a finite and fixed number of solutions to the bulk eigenvector degeneracy points $\mathcal{D}_{ijkl} = 0$ on the full Riemann surface band structure for a general next-nearest-neighbor model. Therefore, the points in Eq.~\eqref{Eq:NNNdeterminant} cannot be created or destroyed under continuous deformations of coupling parameters in Eq.~\eqref{Eq:NNNmodel}. Edge states occur when $\{i, j ,k, l\} = \{1,2,3,4\}$ or $\{i, j ,k, l\} = \{5,6,7,8\}$. The GBZ is given by the 1D curve on the Riemann surface where $|z_4|=|z_5|$. The condition of when the Eq.~\eqref{Eq:NNNdeterminant} solutions correspond to edge states is still related to the $M_i, z_i$ indices. The relabeling of these indices into and away from the edge state condition involves crossing over the GBZ curve, giving the edge state condition a topological origin. The algebraic nature of tight-binding models suggests that the results may be generalized to general longer-range tight-binding models. We leave the question of whether all asymmetric edge states have topological origins in longer-range tight-binding cases open for future work.

The analytical edge state condition for Eq.~\eqref{Eq:analyticalOBC} derived in the End Matter for tight-binding models can be analogously derived for continuous models such as a continuous coupled Hatano-Nelson chain in a periodic potential~\cite{hu2023non}, a $\hat{p}^4$ continuum model with periodic potential~\cite{hu2023non}, one-dimensional magneto-optic photonic crystals~\cite{figotin2001nonreciprocal} under OBC and more. The GBZ may be found from transfer matrix methods~\cite{kunst2019nonhermitian,wielian2025transfer,koekenbier2024transfer,ghaemi2023transport,mong2011edge} or by discretization methods that approximate the continuous model as a tight-binding model~\cite{yokomizo2022nonhermitian,hu2023non}. 
The transfer matrix approaches also have connections to scattering matrix approaches~\cite{fulga2012scattering,peng2017boundary} or boundary Green's function approaches~\cite{guarie2011single,essin2011bulk,mong2011edge,peng2017boundary,fulga2012scattering,thicke2021computing,piasotski2022universal,muller2021universal,rhim2018unified,tamura2021generalization,borgnia2020nonhermitian,boffi2015characterizing,xue2021simple,chen2024formal,hu2023greens}. These studies offer promising directions for extending the study of edge states in continuous models~\cite{xiao2014surface,felbacq2023characterizing} to asymmetric cases. It is possible that the edge states for one-dimensional asymmetric continuous models are topological in nature, analogous to Eq.~\eqref{Eq:restricted} and Eq.~\eqref{Eq:generalinvariant}. Some continuous models may have analytically tractable solutions~\cite{longhi2021nonhermitian,gorlach2019photonic,leykham2017edge,hu2023non,ke2023topological,zhang2020subradiant}. However, for other continuous models such as photonic crystals~\cite{li2025algebraic,yokomizo2022nonhermitian,yokomizo2024nonbloch,figotin2001nonreciprocal,zhong2021nontrivial,yan2021non,ochiai2022non}, the band structure equations are transcendental rather than polynomial, and it becomes difficult to analytically study the bulk eigenvector degeneracies. We leave the problem of whether edge states in these systems are topological as an open question.

\section{Measure of topological robustness}
\label{supp:robustness}
We provide a measure of edge state robustness (or how close an edge state is to being formed) by analyzing how $M_{\mathrm{GBZ}}$ and $M_{\mathrm{deg}}$ respond to perturbations in Eq.~\eqref{Eq:H}. Here, perturbations refer to changes to the couplings or on-site potentials that still obey nearest-neighbor and lattice periodicity (we do not consider robustness to perturbations that break lattice periodicity or the nearest-neighbor assumption). Examples include sub-symmetry-breaking perturbations~\cite{wang2023sub,verma2024non} limited to nearest-neighbor coupling ranges. We present analytical bounds up to first order for Hermitian cases using matrix perturbation theory~\cite{kato2013perturbation} and provide numerical demonstrations for both Hermitian and non-Hermitian models. 

We first derive the first-order correction to the eigenvector ratio for a general non-degenerate matrix under perturbation. Consider 
\begin{equation}
    H=H_0+\delta H.
\end{equation}
where $H_0$ is a $2\times2$ matrix and $\delta H$ is the perturbation on $H_0$. Here, both $H_0$ and $H$ could be non-Hermitian. The two left and right eigenstates of $H_0$ satisfy $\bra{1_L} H_0=\bra{1_L}E^{(1)}$, $\bra{2_L} H_0=\bra{2_L}E^{(2)}$, $H_0 \ket{1_R} = E^{(1)} \ket{1_R}$ and $H_0 \ket{2_R} = E^{(2)} \ket{2_R}$, where $E^{(1)}$ and $E^{(2)}$ are the two eigenvalues of $H_0$. It is possible that $\ket{1_L}\neq \ket{1_R}$ and $\ket{2_L}\neq \ket{2_R}$ for a non-Hermitian $H_0$~\cite{bergholtz2021exceptional,ashida2020nonhermitian,brody2013biorthogonal}.
For later convenience, we do not normalize the eigenvectors but define them in their ratio forms:
\begin{equation}
    |1_R\rangle=\binom{M^{(1)}}{1}, \quad |2_R\rangle=\binom{M^{(2)}}{1}.
\end{equation}
Biorthogonality ($\bra{1_L}\ket{2_R}=\bra{2_L}\ket{1_R}=0$) then allows us to write the left eigenvectors as
\begin{equation}
    \langle 1_L|=\left(\begin{array}{ll}
-1 & M^{(2)}
\end{array}\right), \quad\langle 2_L|=\left(\begin{array}{ll}
-1 & M^{(1)}
\end{array}\right).
\end{equation}
After $H_0$ is perturbed to $H$, the $|1_R\rangle$ eigenvector is perturbed to $|1_R\rangle+|\delta 1_R\rangle$, where 
\begin{align}
|\delta 1_R\rangle &= \delta|1_R\rangle = \binom{\delta M^{(1)}}{0}.
\end{align}
The perturbed vector is an eigenvector of the perturbed matrix, and we can write
\begin{align}
\left(H_0+\delta H\right)(|1_R\rangle+|\delta 1_R\rangle)&=\left(E^{(1)}+\delta E^{(1)}\right)(|1_R\rangle+|\delta 1_R\rangle).
\end{align}
To first order, we have:
\begin{align}
\delta H|1_R\rangle+H_0|\delta 1_R\rangle&=\delta E^{(1)}|1_R\rangle+E^{(1)}|\delta 1_R\rangle \\
\langle 2_L| \delta H|1_R\rangle+E^{(2)}\langle 2_L \mid \delta 1_R\rangle&=E^{(1)}\langle 2_L \mid \delta 1_R\rangle \\
\langle 2_L \mid \delta 1_R\rangle&=\frac{\langle 2_L| \delta H|1_R\rangle}{E^{(1)}-E^{(2)}}.
\end{align}
Using the component forms of the vectors, we have
\begin{align}
    \langle 2_L \mid \delta 1_R\rangle=-\delta M^{(1)}&=\frac{1}{E^{(1)}-E^{(2)}}\left(\begin{array}{cc}
-1 & M^{(1)}
\end{array}\right)\delta H\binom{M^{(1)}}{1}.
    \label{Eq:firstorderMeq}
\end{align}
Applying the Cauchy-Schwarz inequality on the right hand side of Eq.~\eqref{Eq:firstorderMeq} leads to 
\begin{equation}
    \left|\delta M^{(1)}\right| \leq \frac{\left(1+\left|M^{(1)}\right|^2\right)\|\delta H\|}{\left|E^{(1)}-E^{(2)}\right|}.
        \label{Eq:firstorderMbound}
\end{equation}
Here $\|X\|$ is the operator norm of a matrix $X$. For non-Hermitian matrices $X$,  $\|X\|$ is the largest singular value of $X$, which becomes the magnitude of the largest eigenvalue in Hermitian cases. Geometrically, the $M^{(1)}$ point representing $|1_R\rangle$ will move around on the $M$ Riemann sphere. It would be useful to define the chordal distance of two complex numbers on the Riemann sphere, which remains invariant under unitary transformations of $H_0$. Assuming the Riemann sphere has a radius of $1$, the chordal distance between the complex numbers $M^{(1)}$ and $M^{(2)}$ as represented on the $M$-Riemann sphere can be found as
\begin{align}
    d_{\text {chordal}}\left(M^{(1)}, M^{(2)}\right)&=\frac{2\left|M^{(1)} - M^{(2)}\right|}{\sqrt{1+\left|M^{(1)}\right|^2} \sqrt{1+\left|M^{(2)}\right|^2}}.
\end{align}
Applying $d_{\text {chordal}}$ to the change in $M^{(1)}$ leads to, up to first order,
\begin{align}
     d_{\text {chordal, 1st order}}\left(M^{(1)}, M^{(1)}+\delta M^{(1)}\right)&=\frac{2\left|\delta M^{(1)}\right|}{1+\left|M^{(1)}\right|^2}.
\end{align}
Then using Eq.~\eqref{Eq:firstorderMbound}, this first order term in the chordal distance is bounded by
\begin{equation}
    d_{\text {chordal, 1st order}}\left(M^{(1)}, M^{(1)}+\delta M^{(1)}\right) \leq \frac{2\|\delta H\|}{\left|E^{(1)}-E^{(2)}\right|} .
\end{equation}

\begin{figure}[H]
\centering
\includegraphics[width=\textwidth]{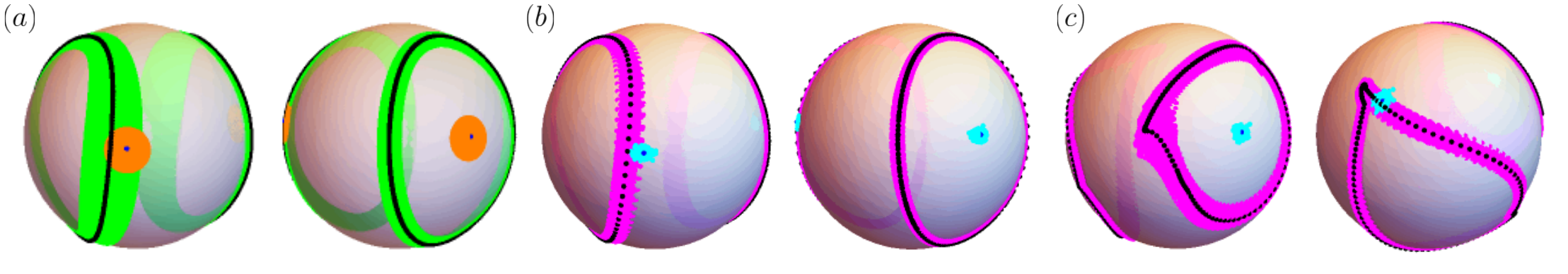}
\caption{(a) First-order perturbation bounds for $H_h+\delta H_h$ for $M_{\text{BZ}}$ in green, $M_{\text{deg}}$ in orange. (b) Numerically solved $H_h+\delta H_h$ for 200 randomly sampled $\delta H_h$ where $\lambda=0.05$. The perturbed $M_{\text{BZ}}$ is in magenta and perturbed $M_{\text{deg}}$ in cyan. (c) Numerically solved $H_{\text{nh}}+\delta H_{\text{nh}}$ for 200 randomly sampled $\delta H_{\text{nh}}$ where $\lambda=0.05$. The perturbed $M_{\text{GBZ}}$ is in magenta and perturbed $M_{\text{deg}}$ in cyan. For (a)-(c) the unperturbed $M_{\text{GBZ}}$ and $M_{\text{deg}}$ are in black and blue respectively.}
\label{fig:perturbation}
\end{figure}

We now apply these perturbation theory results to study the stability of edge states and bulk eigenvector degeneracies for a Hermitian model $H_0=H_h$ of Eq.~\eqref{Eq:H} under a Hermitian perturbation $\delta H=\delta H_h$. The winding number formulation of the number of edge states (Eq. \ref{Eq:restricted}) indicates that an edge state can only be created or destroyed if $M_{\text{deg}}$ moves across $M_{\text{BZ}}^{(1)}$ or $M_{\text{BZ}}^{(2)}$, the image of two Brillouin zone bands $E_{\text{BZ}}^{(1)}$ and $E_{\text{BZ}}^{(2)}$ on the $M$ plane. The stability of such states can be established if a small perturbation leads to finite changes in $M_{\text{deg}}$ and $M_{\text{BZ}}$, which will be shown in the following. As such, the distance between $M_{\text{deg}}$ and $M_{\text{BZ}}$ will also experience finite changes when perturbed and become a measure of overall stability of the existing edge states or the edge states that will be created.

We take the form of perturbation $\delta H_h$ as
\begin{equation}
    \delta H_h = \lambda\left[\begin{array}{cc}
r_a+\frac{s}{z}+s^* z & t+\frac{u}{z}+v z \\
t^*+\frac{v^*}{z}+u^* z & r_b+\frac{w}{z}+w^* z
\end{array}\right]
\label{Eq:dHpert}
\end{equation}
where $r_a,r_b \in [-1,1]$ and $s, t, u, v, w \in \mathbb{D}$ where $\mathbb{D}=\{z \in \mathbb{C} \text{ where } | z | \leq 1\}$ is the unit disk and $\lambda \in \mathbb{R}$ is a scaling parameter. For Hermitian models, $E_{\text{BZ}}$ and $M_{\text{BZ}}$ can be found from Eq.~\eqref{Eq:H} to~\eqref{Eq:M} and using $z=e^{i \theta}$ for $\theta \in [0, 2\pi]$. As such, we can write
\begin{align}
    \delta H_h &=\delta h_{-} z^{-1}+\delta h_0+\delta h_{+} z \\
    \|\delta H_h\| &\leq \left\|\delta h_{-}\right\|+\left\|\delta h_0\right\|+\left\|\delta h_+\right\|
\end{align}
which holds on the BZ. Using Eq.~\eqref{Eq:firstorderMbound}, we then have:
\begin{equation}
    \frac{\left|\delta M^{(1)}_{\text{BZ}}\right|}{1+\left|M^{(1)}_{\text{BZ}}\right|^2} \leq \frac{\left\|\delta h_{-1}\right\|+\left\|\delta h_0\right\|+\left\|\delta h_1\right\|}{\left|E^{(1)}_{\text{BZ}}-E^{(2)}_{\text{BZ}}\right|}.
\end{equation}
As such, the first-order perturbation of $M_{\text{BZ}}$ is bounded on the Riemann sphere when the system is gapped.

For the effects of perturbations on $M_{\text{deg}}$, we observe that Eq. (\ref{Eq:Mdeg}) is equivalent to the condition that $h_+(M_{\text{deg}},1)^T$ and $h_-(M_{\text{deg}},1)^T$ are linearly dependent, where $h_+$ and $h_-$ are the coefficients of $z$ and $z^{-1}$ of $H_0$. As such, $(M_{\text{deg}},1)^T$ is the eigenvector of the following matrix $A$:
\begin{equation}
     A=J h_{-}^T J h_{+}, \quad J=i \sigma_y=\left[\begin{array}{cc}
0 & 1 \\
-1 & 0
\end{array}\right].
\end{equation}
Thus, the perturbation of $M_{\text{deg}}$ can be obtained by applying perturbation theory to $A$. Note that $A$ is non-Hermitian in general, and the two solutions of $M_{\text{deg}}$ do not necessarily form an orthogonal basis. We have
\begin{align}
    \delta A &=J . \delta h_{-}^T . J . h_{+}+J . h_{-}^T . J . \delta h_{+} .
\end{align}
Using Eq.~\eqref{Eq:firstorderMbound}, we then have:
    \begin{align}
    \frac{\left|\delta M_{\text{deg},1}\right|}{1+\left|M_{\text{deg},1}\right|^2} &\leq \frac{\|J . \delta h_{-}^T . J . h_{+}+J . h_{-}^T . J . \delta h_{+} \|}{\left|E_{\text{val},1}-E_{\text{val},2}\right|}.
    \label{Eq:Mdegbound}
\end{align}
where $E_{\text{val},1},E_{\text{val},2}$ are eigenvalues of $A$ and are not the same as $E_{\text{deg}}$ solutions in Eq.~\eqref{Eq:Edeg}. As such, the first-order perturbations of $M_{\text{deg}}$ points are also bounded on the Riemann sphere when $A$ is not degenerate (i.e. $M_{\text{deg}}$ points are distinct).

We have not analytically determined perturbation results for non-Hermitian edge states as replacing the BZ with the GBZ makes the analytical procedures more difficult. Instead, numerical results demonstrate the topological robustness for both Hermitian and non-Hermitian models. Fig.~\ref{fig:perturbation}(a) shows analytical first-order bounds for $H_h$ with Hermitian perturbation ($\lambda=0.05$). Fig.~\ref{fig:perturbation}(b) presents 200 numerical models of $H_h + \delta H_h$ with random parameters, with $M_{\text{BZ}}$ bounds in green/pink and $M_{\text{deg}}$ bounds in orange/cyan respectively. The overlap between these bounds indicates topological stability. If $M_{\text{deg}}$ is an edge state and the bounds overlap, perturbations may cause $M_{\text{deg}}$ to cross over $M_{\text{BZ}}$ and remove the edge state; if the bounds remain separate, the edge state is robust under the perturbations considered. Similarly, when $M_{\text{deg}}$ is not an edge state, overlapping bounds suggest that perturbations may create an edge state. Fig.~\ref{fig:perturbation}(c) demonstrates numerical results for a non-Hermitian model using $H_0=H_{\text{nh}}$ and perturbation $\delta H_{\text{nh}}$ ($\lambda=0.05$), showing $M_{\text{GBZ}}$ and $M_{\text{deg}}$ for 200 models. A more detailed analysis of matrix perturbation may provide tighter bounds or a bound for the entire perturbation instead of the first-order term~\cite{dopico2000weyl,chen2006note,ashida2020nonhermitian,brody2013biorthogonal}.

\section{Numerical details for phase diagrams in Fig.~\ref{fig:phasediagram}}
\label{supp:phasediagram}
\subsection{Hermitian phase diagram}
\label{supp:phasediagramHermitian}

An example of the procedure for obtaining the numerical phase diagram in Fig.~\ref{fig:phasediagram}(a) for the Hermitian model is shown in Fig.~\ref{fig:numericalstepsH} for the model $H_{\text{h}}$. We numerically solve the open boundary spectrum $E_{\text{obc}}$ with $N=32$ and plot it in Fig.~\ref{fig:numericalstepsH}(a). We also solve for $E_{\text{deg}}$ using Eq.~\eqref{Eq:Edeg} (one could also solve Eq.~\eqref{Eq:MzE1} and~\eqref{Eq:MzE2} simultaneously). Using $E_{\text{deg}}$, we can apply the ``analytical edge state condition'' where we classify an $E_{\text{deg}}$ point as an edge state if and only if $|\frac{M_1}{M_2} - 1| < \epsilon$ or $|\frac{M_3}{M_4} - 1| < \epsilon$ at that point for some small $\epsilon$ tolerance (here we choose $\epsilon=0.001$). We plot the $E_{\text{deg}}$ point as a green circle if it corresponds to an edge state, and a red circle otherwise in Fig.~\ref{fig:numericalstepsH}(a). The number of $E_{\text{deg}}$ that are edge states can be counted and is listed in the title as the number of ``analytical edge states'' in Fig.~\ref{fig:numericalstepsH}(a). This reproduces the left panel of Fig.~\ref{fig:phasediagram}(a).

In Fig.~\ref{fig:numericalstepsH}(b,c), we numerically calculate the Hermitian topological invariant in Eq.~\eqref{Eq:restricted}. There are two energy bands in the gapped case, and their image on the $M$-plane is $M(\mathcal{C}_{\text{BZ}}^{(1)})$ and $M(\mathcal{C}_{\text{BZ}}^{(2)})$. For the BZ, we use $z\in e^{i\theta}$ with 500 equally-spaced values of $\theta \in [-\pi,\pi]$. We substitute these $z$ values into $E_\pm(z) = d_0(z) \pm \sqrt{d_x(z)^2 + d_y(z)^2+d_z(z)^2}$. With the $(z,E)$ pairs, we use Eq.~\eqref{Eq:M} to get $M(\mathcal{C}_{\text{BZ}}^{(1)})$ and $M(\mathcal{C}_{\text{BZ}}^{(2)})$. We plot $M(\mathcal{C}_{\text{BZ}}^{(1)})$ and $M(\mathcal{C}_{\text{BZ}}^{(2)})$ \textit{separately} in Fig.~\ref{fig:numericalstepsH}(b) and (c) respectively. We fit smooth curves to $M(\mathcal{C}_{\text{BZ}}^{(1)})$ and $M(\mathcal{C}_{\text{BZ}}^{(2)})$ and use them to numerically count the number of $M_{\text{deg}}$ points inside $M(\mathcal{C}_{\text{BZ}}^{(i)})$. We plot \textit{both} $M_{\text{deg}}$ points (calculated from Eq.~\eqref{Eq:Mdeg} or using $E_{\text{deg}}$) in Fig.~\ref{fig:numericalstepsH}(b) and (c). $W_{\text{BZ}}^{(i)}$ is numerically found by counting the number of $M_{\text{deg}}$ points inside $M(\mathcal{C}_{\text{BZ}}^{(i)})$ and taking mod 2. The direction of winding is not relevant due to the mod 2 and a parity count is sufficient. 

In Fig.~\ref{fig:numericalstepsH}(d) we plot $M(\mathcal{C}_{\text{BZ}})$ and $M_{\text{deg}}$ on the $M$-Riemann sphere in black and blue respectively. 

\begin{figure}[H]
\centering
\includegraphics[width=0.8\textwidth]{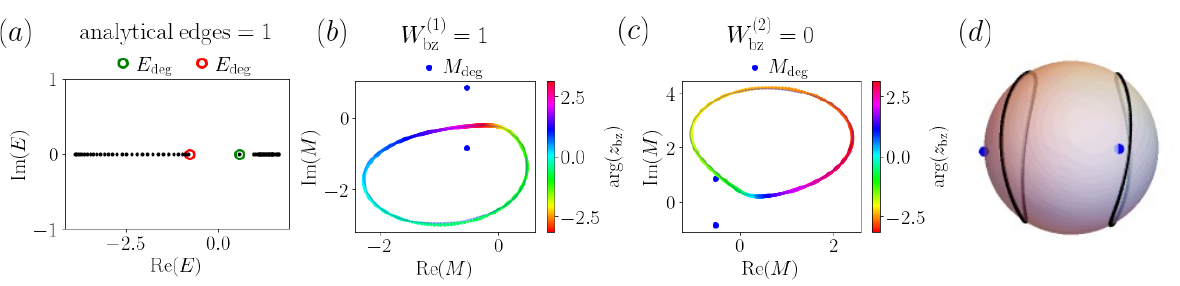}
\caption{(a) $E_{\text{obc}}$ on complex $E$ plane. $E_{\text{deg}}$ is plotted as a green circle if it corresponds to an edge state and a red circle if it does not (with total number of analytical edge states in the title). (b,c) $M(\mathcal{C}_{\text{BZ}}^{{(i)}})$ colored by $\arg(z)$ on the $M$ plane. Both $M_{\text{deg}}$ points are in blue, and the numerically calculated $W_{\text{BZ}}^{(i)}$ from Eq.~\eqref{Eq:restricted} is presented in the title. Here $i=1,2$ is for (b) and (c) respectively. (d) $M(\mathcal{C}_{\text{BZ}})$ and $M_{\text{deg}}$ on the $M$-Riemann sphere in black and blue respectively. All subplots are for the model $H_{\text{h}}$ for $N=32$.}
\label{fig:numericalstepsH}
\end{figure} 

\subsection{Non-Hermitian phase diagram}

An example of the procedure for obtaining the numerical phase diagram in Fig.~\ref{fig:phasediagram}(b) for the non-Hermitian model is shown in Fig.~\ref{fig:numericalstepsNH} for the model $H_{\text{nh}}$. 
The calculation for $E_{\text{obc}}$ and $E_{\text{deg}}$ is the same for the non-Hermitian case as for the Hermitian case described in SM Sec.~\ref{supp:phasediagramHermitian} and is depicted in Fig.~\ref{fig:numericalstepsNH}(a) for the non-Hermitian model. In Fig.~\ref{fig:numericalstepsNH}(b-e), we show the steps to calculate the general invariant in Eq.~\eqref{Eq:generalinvariant} for our non-Hermitian model. Unlike the Hermitian model, we must numerically calculate the GBZ. Different methods are described in Ref.~\cite{wang2024nonhermitian}. For non-Hermitian models, the GBZ topology 
determines whether $M(\mathcal{C}_{\text{GBZ}})$ consists of distinct loops in the $M$-plane, which affects fitting a curve (or multiple curves) to $M(\mathcal{C}_{\text{GBZ}})$. For the phase diagram in Fig.~\ref{fig:phasediagram}(b), we observe that there were two $z_{\text{branch}}$ points enclosed by the GBZ on the $z$-plane for nearly the entire phase diagram. While there are braiding transition within the parameter space of Fig.~\ref{fig:phasediagram}(b), the transition happens over a very small parameter range in comparison to the grid size of the $t_1$ and $\alpha$ values, so we can reasonably approximate the non-Hermitian model as being in an unlinked GBZ phase across the models evaluated in Fig.~\ref{fig:phasediagram}(b). After knowing the braiding topology of the GBZ, we can sort the calculated GBZ into subGBZ~\cite{fu2023anatomy,yang2020nonhermitian,fu2024braiding} loops. Here, the two links of the unlink are the subGBZ loops. An analytical method to sort the subGBZ loops is described in Ref.~\cite{fu2024braiding}, and other clustering algorithms based on machine learning~\cite{shi2024machine,chen2024machine,yu2022experimental,long2024unsupervised,yu2021unsupervised} are also available. In Fig.~\ref{fig:numericalstepsNH}(b,c) we plot the two subGBZ loops on the $z$ plane to ensure that they have been sorted properly.

By sorting the subGBZ loops each by $\arg(z)$, we fit a curve to the distinct $M(\mathcal{C}_{\text{GBZ}}^{(i)})$ loops by sorting $M(\mathcal{C}_{\text{GBZ}}^{(i)})$ by its respective $\arg(z)$ value from the GBZ. In Fig.~\ref{fig:numericalstepsNH}(d,e) we plot \textit{both} $M(\mathcal{C}_{\text{GBZ}}^{(1)})$ and $M(\mathcal{C}_{\text{GBZ}}^{(2)})$ loops but plot $M_{\text{deg},1}$ and $M_{\text{deg},2}$ points \textit{separately} for Fig.~\ref{fig:numericalstepsNH}(d) and Fig.~\ref{fig:numericalstepsNH}(e) respectively. Note that this contrasts to the process in the Hermitian case in SM Sec.~\ref{supp:phasediagramHermitian} and is a consequence of the way Eq.~\eqref{Eq:restricted} and Eq.~\eqref{Eq:generalinvariant} is constructed. $M_{\text{branch}}$ is plotted in red in Fig.~\ref{fig:numericalstepsNH}(d,e). Using the information in Fig.~\ref{fig:numericalstepsNH}(d) and (e) one can calculate $W_1$ and $W_2$ respectively from Eq.~\eqref{Eq:generalinvariant}. The calculation requires finding the winding number of $M_{\text{deg},j}$ (which is $W_{\text{deg},j}$) and the winding number of any one of the four $M_{\text{branch}}$ points (which is $W_{\text{branch}}$). We again calculated the winding numbers by using a count of $M_{\text{deg},j}$ or $M_{\text{branch}}$ in the fitted $M(\mathcal{C}_{\text{GBZ}})$ loops and these counts are listed in the second line of the title of Fig.~\ref{fig:numericalstepsNH}(d,e). For non-Hermitian models, one must in addition check whether there are intersecting $M(\mathcal{C}_{\text{GBZ}})$ loops when numerically counting the winding numbers.

In Fig.~\ref{fig:numericalstepsH}(f) we plot $M(\mathcal{C}_{\text{GBZ}})$, $M_{\text{deg}}$, $M_{\text{branch}}$ on the $M$-Riemann sphere in black, blue and red respectively. 
\begin{figure}[H]
\centering
\includegraphics[width=0.7\textwidth]{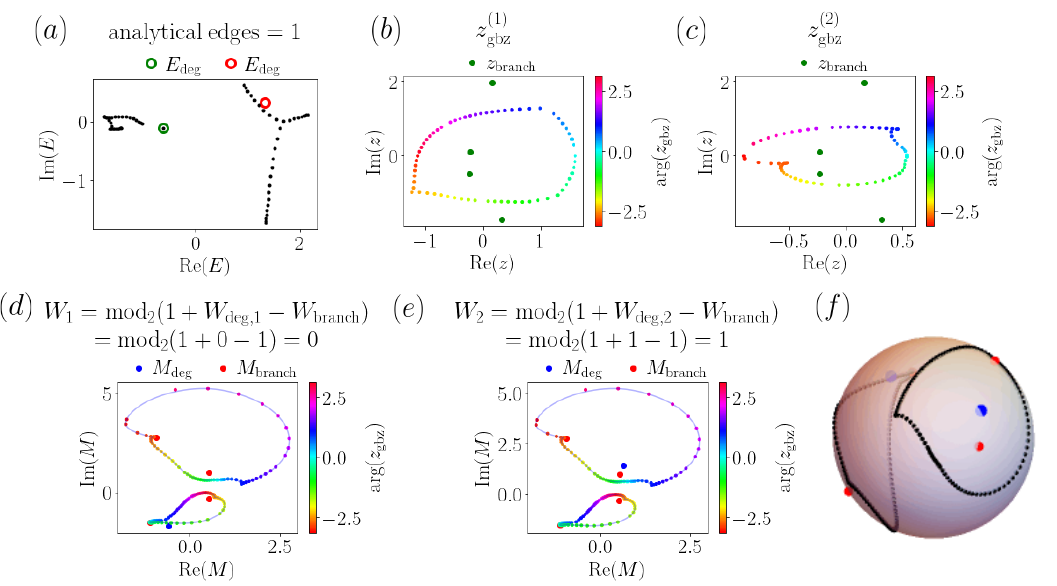}
\caption{(a) $E_{\text{obc}}$ on complex $E$ plane. $E_{\text{deg}}$ is plotted as a green circle if it corresponds to an edge state and a red circle if it does not (with total number of analytical edge states in the title). (b,c) $z$ from $\mathcal{C}_{\text{GBZ}}^{(i)}$ is plotted on the complex $z$ plane colored by $\arg(z)$. $z_{{\text{branch}}}$ is plotted in green. (d,e) $M(\mathcal{C}_{\text{GBZ}})$ colored by $\arg(z)$ on the $M$ plane. $M_{\text{branch}}$ is plotted in red. We plot $M_{\text{deg},j}$ and numerical calculation of $W_{j}$ from Eq.~\eqref{Eq:generalinvariant} is presented in the title for $j=1,2$ in (d) and (e) respectively. (f) $M(\mathcal{C}_{\text{GBZ}})$, $M_{\text{deg}}$ and $M_{\text{branch}}$ on the $M$-Riemann sphere in black, blue and red respectively. All subplots are for the model $H_{\text{nh}}$ for $N=32$.}
\label{fig:numericalstepsNH}
\end{figure}
\subsection{Discussion on the phase diagrams}
\label{supp:numericaldiscussion}
\begin{figure}[H]
\centering
\includegraphics[width=\textwidth]{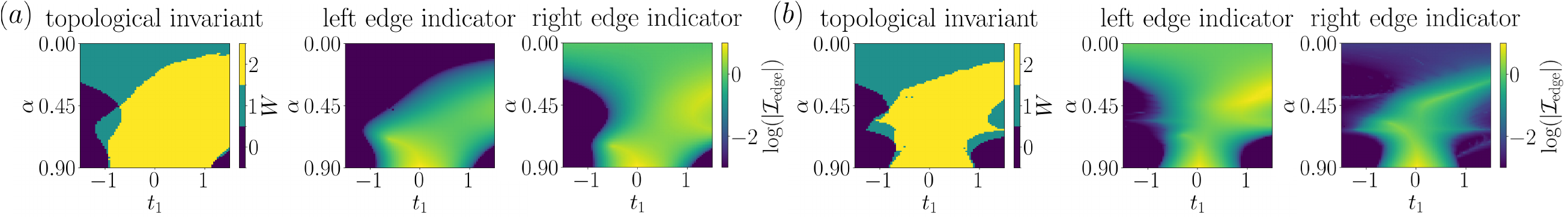}
\caption{Topological invariant and left and right edge indicator phase diagram for Fig.~\ref{fig:phasediagram}.}
\label{fig:phasediagramleftright}
\end{figure}
In the above, the $E_{\text{deg}}$ or $M_{\text{deg}}$ values were evaluated to calculate the topological invariant in Eq.~\eqref{Eq:restricted} and Eq.~\eqref{Eq:generalinvariant}. However, from $E_{\text{deg}}$ or $M_{\text{deg}}$, one could have already found the number of edge states by applying the analytical edge state criteria. Thus, Eq.~\eqref{Eq:restricted} and Eq.~\eqref{Eq:generalinvariant} does not provide a more efficient computational technique to calculate edge states. The significance of Eq.~\eqref{Eq:restricted} and Eq.~\eqref{Eq:generalinvariant} is that it quantifies the topological stability or robustness of the edge or non-edge states calculated from $E_{\text{deg}}$ (see SM Sec.~\ref{supp:robustness}). We also note that Fig.~\ref{fig:phasediagram}(a) and (b) have similar overall shapes. This is because we deliberately chose randomly generated models for $H_{\text{h}}$ and $H_{\text{nh}}$ that both had only one edge state, as they showed more features on the phase diagram for the parameters in Fig.~\ref{fig:phasediagram}. Thus, the plots match at $\alpha=0$ and $\alpha=1$ but are otherwise unrelated.

We also briefly discuss some alternative representations of the phase diagrams in Fig.~\ref{fig:phasediagram}. Instead of using $E_{\text{deg}}$ for the analytical edge state criteria, one could also find $M_1$ to $M_4$ for $E$ of all OBC modes of a finite model and similarly apply the $|\frac{M_1}{M_2} - 1| < \epsilon$, $|\frac{M_3}{M_4} - 1| < \epsilon$ criteria. For Hermitian models, localization indicators~\cite{ganeshan2015nearest} such as the IPR or wavefunction center-of-mass can also detect edge states, and these could be used to compare OBC mode properties to the topological invariants. For non-Hermitian systems, these indicators are more complicated due to the possibility of localization properties changing with imaginary gauge transforms and bulk and edge states having similar localization lengths. One could use a `gauged IPR' computed from eigenvectors transformed as $(z_2^{-n}\psi_{na}, z_2^{-n}\psi_{nb})$ and $(z_3^{-n}\psi_{na}, z_3^{-n}\psi_{nb})$. The $(z_2^{-n}\psi_{na}, z_2^{-n}\psi_{nb})$ eigenvectors would have bulk and $M_1=M_2$ edge states be mainly plane waves and $M_3=M_4$ edge states would be localized using the leading term analysis in SM Sec.~\ref{supp:modeprofiles}. Similarly, $(z_3^{-n}\psi_{na}, z_3^{-n}\psi_{nb})$ eigenvectors would have bulk and $M_3=M_4$ edge states be mainly plane waves and $M_1=M_2$ edge states would be localized. However, when there are two edge states and they are close in energy, the leading terms in the mode profiles can change and the edge states can be hybridized modes. 

Another edge-state indicator applicable to our non-Hermitian models is the signed quantity $\mathcal{I}_{\text {edge }}= \pm \ln \left(\left|z_3 / z_2\right|\right)$, where we assign a negative sign if $\left|\frac{M_1}{M_2}-1\right|<\left|\frac{M_3}{M_4}-1\right|$ and a positive sign otherwise. $\mathcal{I}_{\text {edge }}$ vanishes for bulk states (since $\left|z_2\right|=\left|z_3\right|$) and is finite for edge states $\left(\left|z_2\right| \neq\left|z_3\right|\right)$. The edge indicator detects an $M_1=M_2$ edge state if it is negative and finite and an $M_3=M_4$ edge state if it is positive and finite. The magnitude of $\mathcal{I}_{\text{edge}}$ is invariant under gauge transforms to the model. From SM Sec.~\ref{supp:generalinvariant_DS}, there can be at most one $M_1=M_2$ and one $M_3=M_4$ state for a model in Eq.~\eqref{Eq:H}. Then the minimum and maximum $\mathcal{I}_{\text{edge}}$ values of all OBC modes of a finite model are the $M_1=M_2$ (left) and $M_3=M_4$ (right) edge indicators, respectively, for an Eq.~\eqref{Eq:H} model. We plot these indicators for the parameters in Fig.~\ref{fig:phasediagram} in Fig.~\ref{fig:phasediagramleftright}. Finally, the $E$ of OBC modes with finite $N$ was used to calculate the GBZ in Fig.~\ref{fig:phasediagram}. The GBZ is used to numerically calculate the topological invariants in Fig.~\ref{fig:phasediagram} and so the topological invariants will also have some finite-sized effects. However, there are other ways to obtain the GBZ~\cite{wang2024nonhermitian} with can offer higher accuracy. We also include an animation of the plots in Fig.~\ref{fig:numericalstepsH} for a slice of the Hermitian phase diagram along $t_1=1.5$ verifying Eq.~\eqref{Eq:restricted} in Fig.~\ref{fig:phasediagram}(a) and an animation of the plots in Fig.~\ref{fig:numericalstepsNH} for the a slice of the non-Hermitian phase diagram along $t_1=-1.01$ verifying Eq.~\eqref{Eq:generalinvariant} in Fig.~\ref{fig:phasediagram}(b) as Supplementary Movies 1 and 2 respectively.

\section{Visualization of the $M$-invariant as an index tracker}
\label{supp:visualM}
In the main text, we stated that the edge-state invariants are index trackers used to determine whether $M_{\text{deg}}$ corresponds to the edge-state condition. We now discuss two different visualizations of this argument. In this section, we will use the following model:
\begin{equation}
    H=(1-\alpha) H_{\text {nh2 }}+\alpha H_{\mathrm{ssh}}
    \label{Eq:Hnh2}
\end{equation}
where $H_{\text {nh2 }}$ is given by:
\begin{table}[H]
\centering
\begin{tabular}{|c|c|c|c|c|} 
\hline 
Model & $h_{-}$ & $h_0$ & $h_{+}$ \\ 
\hline
$H_{\text {nh2 }}$ & 
{$\left[\begin{array}{cc}
0.51+0.40i & -0.44+0.38i \\
-0.86+0.40i & -0.86-0.03i
\end{array}\right]$} & 
{$\left[\begin{array}{cc}
-0.18+0.32i & -0.33+0.70i \\
-0.44-0.84i & 0.21-0.74i
\end{array}\right]$} & 
{$\left[\begin{array}{cc}
0.74-0.13i & -0.43-0.08i \\
-0.79-0.84i & 0.01+0.60i
\end{array}\right]$} \\ 
\hline 
\end{tabular}
\end{table}
\subsection{Edge state transition on $M(E)$ sheets and winding on $M$-plane}
While the collection of $(M,E)$ points does not form a Riemann surface due to the degeneracy, we can still obtain four $M$ values at each $E$ and thus define four $M(E)$ sheets. The bulk eigenvector degeneracies occur when the $M(E)$ sheets touch and show up as minima in $\log(|M_i-M_j|)$, where $i\neq j$ and the $i,j$ labels are derived from the $z$ solution orderings.
The $E$-plane branch points will also show up as minima, but these are expected to be less prominent than the bulk eigenvector degeneracies (due to the square-root behavior near the branch points). This gives an alternative numerical method for locating $E_{\text{deg}}$ and $M_{\text{deg}}$, in contrast to the analytical methods described in SM Sec.~\ref{supp:MdegEdeg}. Note that this is also useful for longer range cases (SM Sec.~\ref{supp:secondnearest}) where analytical solutions may be intractable.

In Fig.~\ref{fig:Mwinding2edge}-\ref{fig:Mwinding0edge}, the first panel shows the minimum $\log(|M_i-M_j|)$ heatmaps with analytically determined $E_{\text{deg}}$ (green if edge state, red if not), $E$-plane branch points (blue), and $E_{\text{obc}}$ (small black points). The second panel displays $E_{\text{obc}}$ in thicker black without the heatmap for clarity. The third panel shows the $z(E)$ sheets with OBC solutions (black). At $E_{\text{deg}}$ there are four $z_1(E)$ to $z_4(E)$ from top to bottom. We only show the two $z$ values that correspond to $M_{\text{deg}}$ (green for edge states, red otherwise). The fourth panel displays $M$-plane images of $\mathcal{C}_{\text{GBZ}}$ colored by $\arg(z)$, with $M_{\text{deg}}$ and $M_{\text{branch}}$ plotted to calculate the edge state invariant.

From Fig.~\ref{fig:Mwinding2edge} to \ref{fig:Mwinding0edge}, the number of edge states in the models decrease from 2 to 0. The $E_{\text{deg}}$ point must cross the $E_{\text{GBZ}}$ curves to change the corresponding $M_i$ indices. This crossing relates to $M_{\text{deg}}$ moving in or out of the $M(\mathcal{C}_{\text{GBZ}})$ loops. The distance between $E_{\text{deg}}$ and $E_{\text{GBZ}}$ indicates how far $M_{\text{deg}}$ is from the $M(\mathcal{C}_{\text{GBZ}})$ loop boundaries, measuring the robustness of edge state formation or removal (see SM Sec.~\ref{supp:robustness}).

\begin{figure}[H]
\centering
\includegraphics[width=\textwidth]{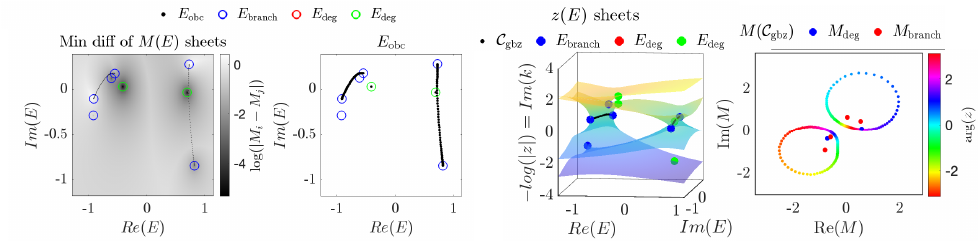}
\caption{Two edge states, $\alpha = 0.5,t_1=1, t_2 = 0.7$ for Eq.~\eqref{Eq:Hnh2}. Here, the two $E_{\text{deg}}$ points correspond to $M_1=M_2$, $M_3=M_4$.}
\label{fig:Mwinding2edge}
\end{figure}

\begin{figure}[H]
\centering
\includegraphics[width=\textwidth]{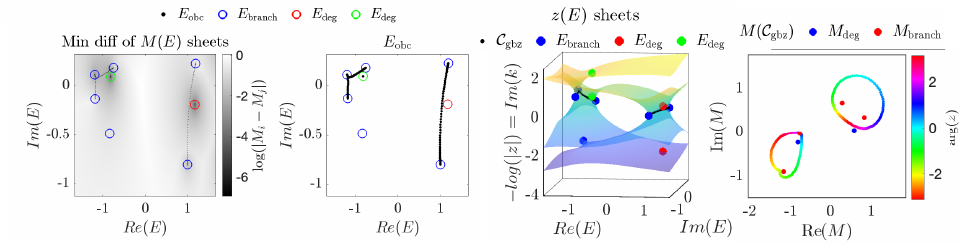}
\caption{One edge state, $\alpha = 0.5,t_1=2, t_2 = 0.9$ for Eq.~\eqref{Eq:Hnh2}. Here, the two $E_{\text{deg}}$ points correspond to $M_1=M_2$, $M_2=M_4$.}
\label{fig:Mwinding1edge}
\end{figure}

 \begin{figure}[H]
\centering
\includegraphics[width=\textwidth]{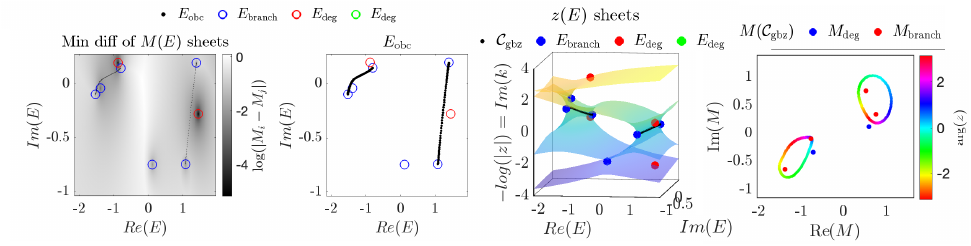}
\caption{No edge states, $\alpha  = 0.5,t_1=2.5, t_2 = 0.7$ for Eq.~\eqref{Eq:Hnh2}. Here, the two $E_{\text{deg}}$ points correspond to $M_1=M_3$, $M_2=M_4$.}
\label{fig:Mwinding0edge}
\end{figure}

\subsection{Auxiliary GBZ on the $M$-Riemann sphere}
Another visualization of Eq.~\eqref{Eq:restricted} and Eq.~\eqref{Eq:generalinvariant} is to map the $M_i$ regions to the $M$-Riemann sphere, where different regions containing $M_{\text{deg}}$ indicate indices of the $M_i=M_j$ condition. These regions can be found by sampling $E$ values, solving for $z_i(E)$, and mapping to $M_i(E)$ via Eq.~\eqref{Eq:M}. The boundaries between regions are images of the auxiliary GBZ~\cite{yang2020nonhermitian} which are the solutions to Eq.~\eqref{Eq:charpoly} where $|z_1|=|z_2|$, $|z_2|=|z_3|$, or $|z_3|=|z_4|$.

Fig.~\ref{fig:AuxGBZTop} shows Eq.~\eqref{Eq:Hnh2} with $\alpha=0.9$ and $t_1=0.5, t_2=1$, containing two edge states. We plot $M_{12}$, $M_{23}$, and $M_{34}$ branches (auxiliary GBZ images) with $M_{\text{deg}}$ points (black markers) and $M_1$ to $M_4$ regions (red, yellow, blue, purple). Some of the $M_i$ regions cover parts of the $M$-Riemann sphere twice, and the colors are more saturated in such regions. One $M_{\text{deg}}$ point lies in $M_1$ and $M_2$ regions and the other in $M_3$ and $M_4$ regions, satisfying edge state conditions. Fig.~\ref{fig:AuxGBZTrivial} uses Eq.~\eqref{Eq:Hnh2} with $\alpha=0.9$ and $t_1=1.5, t_2=1$. Here, one $M_{\text{deg}}$ points lies in $M_1$ and $M_3$ regions and the other point in $M_2$ and $M_4$ regions, conditions that do not produce edge states.

 \begin{figure}[H]
\centering
\includegraphics[width=\textwidth]{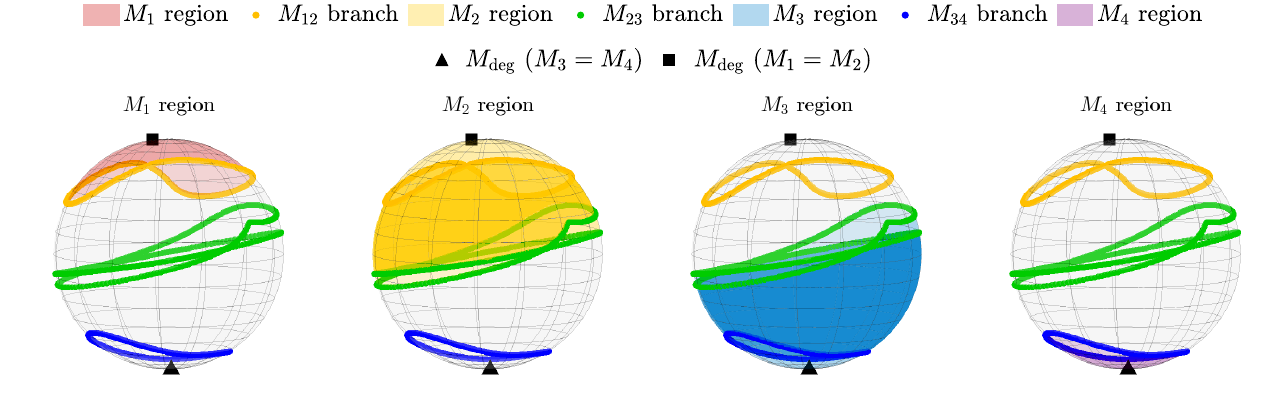}
\caption{The two $M_{\text{deg}}$ points lie in the $M_1,M_2$ regions and the $M_3,M_4$ regions, corresponding to two edge states. Here, $\alpha =0.9, t_1=0.5,t_2=1$ for the model in Eq.~\eqref{Eq:Hnh2}.}
\label{fig:AuxGBZTop}
\end{figure}

 \begin{figure}[H]
\centering
\includegraphics[width=\textwidth]{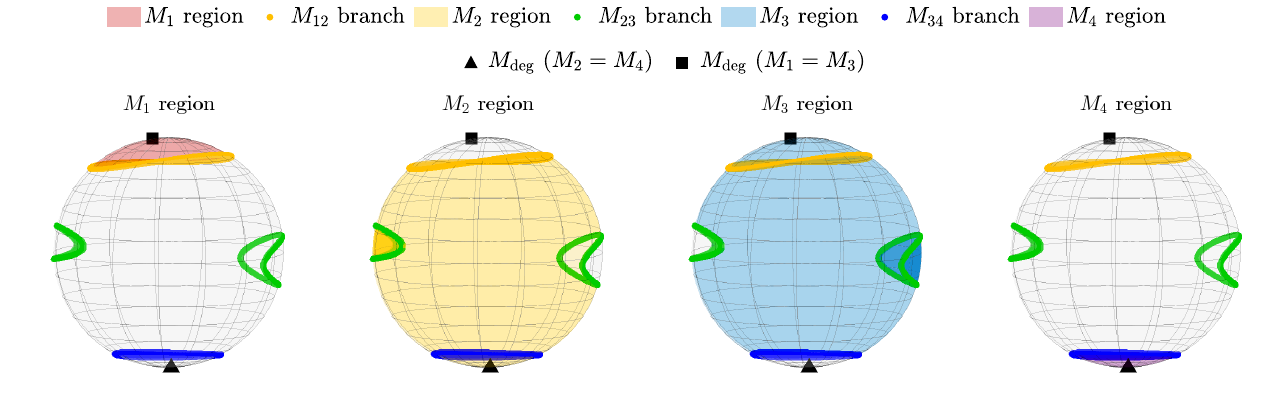}
\caption{The two $M_{\text{deg}}$ points lie in the $M_1,M_3$ regions and the $M_2,M_4$ regions, corresponding to no edge states. Here, $\alpha =0.9, t_1=1.5,t_2=1$ for the model in Eq.~\eqref{Eq:Hnh2}.}
\label{fig:AuxGBZTrivial}
\end{figure}


\end{document}